\documentclass[preprint,aps,singlecolumn,superscriptaddress,showpacs,floatfix,amssymb,amsmath]{revtex4-2}

\usepackage{amsmath,amssymb,bm}
\usepackage{mathptmx}
\usepackage[T1]{fontenc}
\usepackage{microtype}
\usepackage{graphicx}
\usepackage[dvipsnames]{xcolor}
\usepackage{booktabs}
\usepackage[colorlinks=true,allcolors=MidnightBlue]{hyperref}
\usepackage{lineno}

\usepackage{etoolbox}
\AtBeginEnvironment{figure}{\selectfont}
\AtBeginEnvironment{figure*}{\selectfont}

\hypersetup{
  pdftitle={Magnetic flux ring splitting near a \\ Navier-Stokes singularity},
  pdfauthor={Antti J. Niemi}
}

\newcommand{\eB}{\varepsilon_B}

\newcommand{\includefigure}[2][\textwidth]{%
  \IfFileExists{#2}{\includegraphics[width=#1]{#2}}{%
    \fbox{\parbox[c][5cm][c]{\dimexpr#1-2\fboxsep-2\fboxrule\relax}{\centering Figure placeholder\par\texttt{\detokenize{#2}}}}}}

\begin{document}

\title{
\Large%
Magnetic flux ring splitting near a Navier–Stokes singularity
}

\author{Antti J. Niemi}
\email{Antti.Niemi@su.se}
\affiliation{Nordita, Stockholm University,
Hannes Alfv\'ens v\"ag 12, SE-106 91 Stockholm, Sweden}
\affiliation{Wilczek Quantum Center,  Shanghai Institute for Advanced Studies, 
			University of Science and Technology of China, Shanghai 201315, China}

\raggedbottom



\begin{abstract}
OpenAI reports a forced Navier–Stokes flow that becomes singular in finite time, raising the question of its physical consequences. 
Magnetic reconnection offers a test through the splitting and merging of flux structures. We use resistive magnetohydrodynamics 
to simulate a magnetic ring driven by an imposed, regularized flow that reproduces the geometry and scaling of the collapsing core. 
At the nominal singular time, we remove the imposed drive and allow the system to evolve freely. Reconnections before and after 
release split the ring into two or more daughter rings through bifurcations of the poloidal flux function, with Lorentz feedback
selecting the bifurcation type. Unequal winding by the swirl above and below the midplane produces asymmetric toroidal 
magnetic stresses, unfolding the kinematic pitchfork-like transition into a fold. For single-island splits, the cusp normal form 
predicts the asymmetry of the daughter rings without adjustable parameters.  Three-dimensional simulations retain the splits and
merger, confirming that approach to  hydrodynamic blow-up can trigger magnetic restructuring.
\end{abstract}

\maketitle

\section*{Introduction}

Seen in the extreme ultraviolet, the outer atmosphere of the Sun is a thicket of glowing arches. These coronal loops trace tubes
of magnetic flux anchored in the photosphere, where convection shuffles their footpoints without pause\cite{reale2014}. At the
conductivity of the corona the field is approximately frozen into the plasma, and loops that are pushed together deform. Occasionally,
across a thin sheet of current, their field lines break and rejoin, the connectivity changes, and the stored energy is
released in a flare or a coronal mass ejection\cite{priest2000}. Such reconnections can turn an arcade of loops into a
flux rope\cite{vanballegooijen1989}, make colliding flux tubes merge or slingshot apart\cite{linton2001}, and
accompany the eruption of a rope that has lost equilibrium\cite{forbes1991}. The mechanism proposed to 
explain solar flares some seventy years ago\cite{sweet1958,parker1957} has since been observed in the corona\cite{su2013}, in Earth's
magnetosphere\cite{burch2016} and in laboratory plasmas\cite{yamada2010}, and the same process powers some of the most
energetic events in astrophysics\cite{zweibel2009}.

Magnetic helicity measures the linking and twisting of magnetic field lines\cite{moffatt1969}. At high conductivity, it can remain 
nearly conserved during reconnection even as the topology of flux surfaces changes\cite{taylor1974,berger1984,berger1984b}.
In an axisymmetric field these surfaces are level sets of a flux function $\Psi(r, z)$.
Local changes in their topology can be tracked through bifurcations of its critical points, the magnetic axes and saddles that appear
as O and X points in the poloidal plane. The tearing instability creates O--X pairs in a current sheet\cite{furth1963,loureiro2007},
coalescence annihilates them\cite{finn1977}, the tokamak sawtooth exchanges one magnetic axis for another\cite{kadomtsev1975},
and internal reconnection can divide a coronal flux rope in a partial eruption\cite{gibson2006}.

In three dimensions the  changes of topology have been classified through bifurcations of the null points and separatrix surfaces of the 
field\cite{hornig1996,priest1996,priest1997,brown1999}, 
with the caveat that reconnection without a null point changes connectivity in ways the magnetic skeleton alone does not 
show\cite{schindler1988,priest1995}. The classifications establish which transitions are possible, but do not answer which are realized, 
when, or in what order.  Here we address these dynamical questions by following the splitting of a magnetic ring driven 
by an imposed flow. Which bifurcations occur, and when, depends on the competition between the strain that deforms the field, 
the Lorentz force that resists it, and magnetic diffusion, which allows the topology to change.

With its strain growing without bound, a flow approaching a finite-time singularity offers a natural test of this
competition. A recent manuscript from OpenAI reports such a flow\cite{openai2026} as  a solution of the
three-dimensional incompressible Navier--Stokes equations under a smooth body force. The solution describes
an axisymmetric core collapse, with fluid spiralling inward towards a stagnation point and
expelled along the axis, and length scales shrinking to zero at the singular time.
Whether an initially smooth viscous flow can develop such a singularity is a longstanding question 
dating back to Leray\cite{leray1934}. He envisaged a self-similar collapse in which the kinetic 
energy of the shrinking core vanishes at a finite singular time, even though the velocity diverges; 
the reported flow shares this property\cite{openai2026}. 
Other settings also exist that illustrate how singular behaviour can 
accompany changes in shape and connectivity\cite{eggers2015,moffatt2019}. For example, drop pinch-off separates 
one body of fluid into two through a transient finite-time singularity of the
Navier--Stokes equations with a free surface\cite{eggers1997}. In ideal magnetohydrodynamics, currents 
concentrate into sheets of vanishing thickness, setting the stage for reconnection\cite{parker1972}. 
We therefore examine what observable changes the flow reported by OpenAI might trigger. For this we
place  a magnetic flux ring in an imposed flow that reproduces the geometry and self-similar scaling of the
collapsing core\cite{openai2026,leiren2026}. This flow compresses the ring until the singular time, 
at which the unregularized core has lost its kinetic energy, and following Leray\cite{leray1934} we then remove 
the sustaining force and release the system.

Our initial ring is a Taylor spheromak with linked toroidal and 
poloidal flux\cite{taylor1986}. We solve the resistive magnetohydrodynamic equations 
across three decades in magnetic coupling at two magnetic diffusivities. 
We find that magnetic feedback controls both when and how the ring splits. For a rounded torus, a pitchfork-like split at 
weak coupling gives way to a fold at stronger coupling. A cusp catastrophe organizes both, with an asymmetry 
parameter that we measure from the evolving magnetic geometry.
For single-island splits, the cusp normal form predicts the island-depth ratio from the positions of the axes and saddle, in agreement with the ratio
measured from the simulated flux. After release, the flow left by the collapse can split the daughter rings again 
and, at strong coupling, merge their inner cores into an additional
central ring, producing a triplet. Compact tori compressed in the laboratory have comparable parameters and offer
a setting in which to test the predicted sequence of splitting and merging.

\section*{Results}

\subsection*{The equations and the collapse background}

We base the imposed collapsing flow on the reconstruction\cite{leiren2026} of the leading-order profile of the
reported solution\cite{openai2026}. The profile is axisymmetric, with a swirl about the axis. The singular time is
$t = 1$, and at $t = 0$ we place our axisymmetric magnetic ring in the flow, with its twist parallel to the winding
by the swirl above the midplane. The imposed flow and the response are therefore solved in axisymmetry, with two
representative cases repeated in full three dimensions to test for non-axisymmetric instabilities.
The fluid and the magnetic field $\mathbf B$ evolve according to the equations of incompressible resistive
magnetohydrodynamics\cite{alfven1942,batchelor1950}. Taking the seed ring radius as the unit of length and the 
collapse time as the unit of time, the equations are
\begin{align}
\partial_t\mathbf u + (\mathbf u\cdot\nabla)\mathbf u &= -\nabla p + \nu\nabla^2\mathbf u
+ \eB\,(\nabla\times\mathbf B)\times\mathbf B + \mathbf f, \nonumber\\
\nabla\cdot\mathbf u &= 0, \label{eq:ns}\\
\partial_t\mathbf B &= \nabla\times(\mathbf u\times\mathbf B) + \eta\nabla^2\mathbf B, \qquad
\nabla\cdot\mathbf B = 0, \label{eq:ind}
\end{align}
where $\nu$ and $\eta$ are the dimensionless kinematic viscosity and magnetic diffusivity, respectively, and $\eB$ is
the magnetic coupling, which sets how strongly the field acts back on the flow. Equation (\ref{eq:ns}) reduces to the
forced Navier--Stokes equations when $\mathbf B = 0$, and the reported singular flow\cite{openai2026} is a solution of this 
equation under a smooth body force $\mathbf f$.

We write the total velocity as $\mathbf u=\mathbf U+\mathbf v$, where $\mathbf U(r, z, t)$ is the prescribed incompressible
collapse flow and $\mathbf v$ is the dynamical response. The force is chosen so that $\mathbf U$ alone is an exact solution
of equation (\ref{eq:ns}) with $\mathbf B = 0$ and $\mathbf v = 0$, which gives
$\mathbf f = \partial_t\mathbf U + (\mathbf U\cdot\nabla)\mathbf U - \nu\nabla^2\mathbf U + \nabla p_U$, with 
$p_U$ is the pressure of the imposed flow  (Methods).  Equations (\ref{eq:ns}) and (\ref{eq:ind}) are then solved for $\mathbf v$ and $\mathbf B$, with
$\mathbf v(0) = 0$, in the background of the collapse. The field is advected and stretched by $\mathbf u = \mathbf U + \mathbf v$,
the Lorentz force drives $\mathbf v$, and $\mathbf v$ feeds back on $\mathbf B$, while $\mathbf U$ itself is never modified
by the field.

In the unregularized profile the singular core shrinks radially like $\sqrt{1-t}$ and axially like
$(1-t)^{(1-\delta)/2}$, so fast that its kinetic energy vanishes like $(1-t)^{(1-3\delta)/2}$, with $\delta = 0.005$,
while its velocity diverges (Methods). This concerns the shrinking core alone. The surrounding flow need not vanish,
and the swirl of our imposed profile extends beyond the core. In our simulations the collapse is regularized on a short 
time scale $\tau_f = 0.05$. The strain, which is the rate
at which the inflow compresses the ring radially and stretches it along the axis, then grows about twentyfold instead of
without bound. The meridional flow is confined to a column around the axis that contracts with the collapse (Methods).

Leray's hypothetical self-similar collapse loses its kinetic energy and continues after the singular time as the state
of rest\cite{leray1934}. Although with finite energy it needs a force before that time\cite{nrs1996,tsai1998}, the
state of rest that follows needs none. Following this, we release the regularized state at $t=1$ by transferring the
imposed velocity to the dynamical response and removing the sustaining force. The total velocity, magnetic field,
energies and helicity remain continuous across $t=1$, while the acceleration jumps by the solenoidal part of that
force (Methods). The state produced by the collapse relaxes, as would a magnetized ring released from compression in
the laboratory. We adopt the release as a model choice, without claiming a continuation of the singular construction,
and state where the results depend on it. At $\tau_f = 0.05$ the imposed flow at the  release still carries 43\% of its  initial
poloidal energy, and this fraction scales as $\tau_f^{1/2}$ (Supplementary Note 4). At $\eB = 1000$, $\eta = 0.05$, the
one case tested, a smooth switch-off of the force over 0.05 preserves the sequence of events after the release but
delays them by up to about 0.4 and adds one transient pair (Supplementary Note 5).

\begin{figure*}[tbp]
\centering
\includefigure[\textwidth]{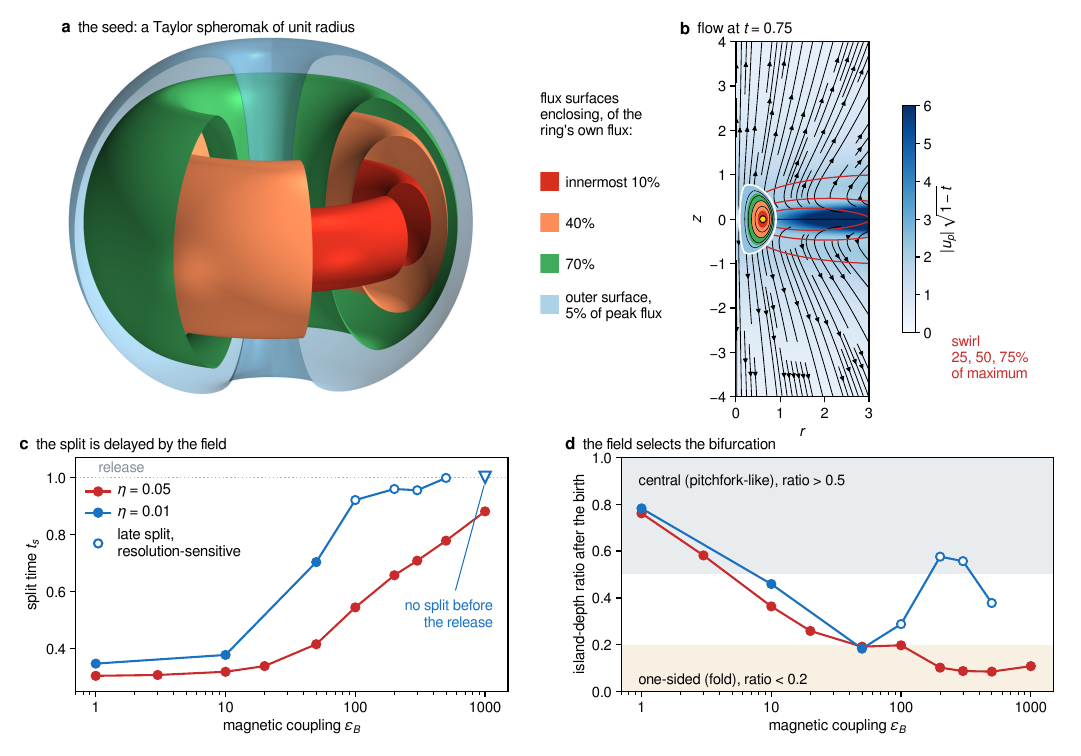}
\caption{\textbf{The torus, the collapse flow and the splitting.}
\textbf{Panel a}, the initial torus, a Taylor spheromak of unit radius, in the colour code of every ring in this work:
flux surfaces enclosing 10, 40 and 70\% of the ring's poloidal flux (red, orange, green) and
the outer surface at 5\% of the peak flux (blue), revolved about the axis and cut open.
\textbf{Panel b}, the imposed flow at $t = 0.75$: speed of the meridional flow times $\sqrt{1-t}$ (blue), streamlines
(black), swirl at 25, 50 and 75\% of its maximum (red) and the initial torus as in Panel a, with flux
contours and its magnetic axis (gold). Fluid is drawn in along the midplane and expelled along the axis inside a
column contracting like $\sqrt{1-t}$.
\textbf{Panel c}, split time, the birth of the first island on the flux ridge that grows deeper than
$3\times10^{-3}\Psi_{\max}$, against the magnetic coupling $\eB$. At $\eta = 0.05$ it rises from $t = 0.31$ at
$\eB = 1$ to 0.88 at $\eB = 1000$. At $\eta = 0.01$ it jumps from 0.70 to 0.92 between $\eB = 50$ and 100, and the
four late splits ($\eB = 100$--500) fall at $t = 0.92$--1.00. At $\eB = 1000$, $\eta = 0.01$ no split occurs before
the release (triangle).
\textbf{Panel d}, type of the split, the ratio of the island depths on the two sides of the new saddle when the new
island first exceeds $10^{-3}\Psi_{\max}$, near 1 for a central split (pitchfork) and near 0 for a one-sided one
(fold). Shaded bands mark the ratios above 0.5 and below 0.2. At $\eta = 0.05$ the ratio falls from 0.76 at $\eB = 1$
to 0.19--0.20 at $\eB = 50$--100 and about 0.1 beyond, crossing 0.5 near $\eB = 4$ and 0.2 near 40. At $\eta = 0.01$
it falls from 0.78 to 0.18 at $\eB = 50$ and returns to 0.56--0.58 at $\eB = 200$--300. Open circles in
Panels c and d, the four late splits at $\eta = 0.01$, whose timing and symmetry are
resolution-sensitive (Methods).}
\label{fig1}
\end{figure*}

Figure~\ref{fig1}a shows the initial torus, a Taylor spheromak of unit radius, and Fig.~\ref{fig1}b the imposed
flow at $t = 0.75$ with the torus in its path (Methods). At unit field the Alfv\'en time across the torus is $1/\sqrt{\eB}$,
which equals the collapse time at $\eB = 1$ and a thirtieth of it at $\eB = 1000$. The magnetic diffusivity is $\eta = 0.05$ or $0.01$ and the
viscosity $\nu = 0.1$ in the same units. At the release the Lundquist number lies between 
5 and 4300, the Reynolds number of the flow is about 500 and the magnetic Prandtl number is 2 or 10 (Methods). 
Compressed and merging compact tori in the laboratory have comparable values\cite{brown2015,lindemuth1983}.

%
%
%
%
%
%
%
%
%

%
%
%
%
%
%
%
%
%
%
%
%

%
%
%
%
%
%
%
%
%
%
%
%

\begin{figure*}[tbp]
\centering
\includefigure[\textwidth]{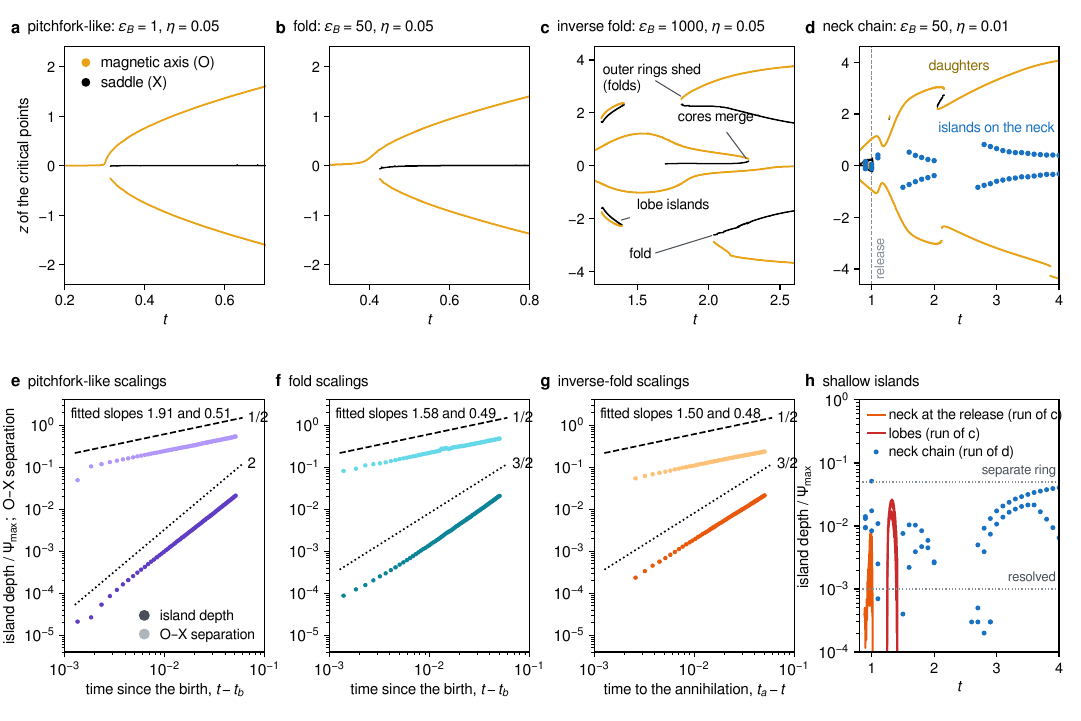}
\caption{\textbf{Splitting, merging and shallow islands on the flux ridge.} World lines of the resolved magnetic
axes (gold) and saddles (black) along the flux ridge, and the normal-form scalings of the same events.
\textbf{Panel a}, pitchfork-like split at $\eB = 1$, $\eta = 0.05$. The original axis moves slightly off $z = 0$ while a
saddle and a second, nearly equal axis appear beside it.
\textbf{Panel b}, fold at $\eB = 50$, $\eta = 0.05$. The saddle and the new axis are born together off the stagnation
plane, and the new island is born much shallower.
\textbf{Panel c}, inverse fold at $\eB = 1000$, $\eta = 0.05$, after the release. Shallow islands form on the lobes of
both daughters at $1.25 < t < 1.41$, the daughters shed outer rings in folds at $t = 1.81$ and $2.03$, and their inner
cores merge at $t = 2.28$ into the central ring of the triplet.
\textbf{Panel d}, island chain on the neck at $\eB = 50$, $\eta = 0.01$. Blue, the islands on the neck deeper than
$10^{-3}\Psi_{\max}$ in the ring records, every 0.1 time units. The jumps of the daughters' axes near $t = 2.1$ are a
fold and an inverse fold inside each daughter.
\textbf{Panels e}--\textbf{g}, island depth (dark) and axis--saddle separation (light) against the time since the birth
(\textbf{e}, \textbf{f}) or to the annihilation (\textbf{g}) for the events of Panels a--c, with the
reference slopes 2, 3/2 and 1/2. The birth and annihilation times, $t_b = 0.304$ and $0.418$ and $t_a = 2.282$, are
fitted from the separation (Methods).
\textbf{Panel h}, depth histories of the shallow islands. The neck island at the release and the lobe islands of
Panel c are transient, with maximum depths of 0.008 and 0.026, while the neck chain of Panel d stays
between $10^{-3}$ and $5\times10^{-2}$ to the end of the run. Dotted lines, the resolution threshold $10^{-3}$ and the
5\% threshold of a separate ring.}
\label{fig2}
\end{figure*}

\subsection*{Field strength selects the bifurcation}

The flux surfaces drawn in Fig.~\ref{fig1}a are level sets of the flux function $\Psi(r, z)$. The torus comes
apart when $\Psi$ acquires a second maximum, corresponding to a second magnetic axis. Extrema of $\Psi$ are the O points of the
poloidal field and saddles are its X points. With $\Psi$ smooth and no critical point crossing the boundary of the
tracked region, the Poincar\'e--Hopf theorem fixes the number of O points minus the number of X points. Every change 
in the number of critical points is then the creation or annihilation of an O--X pair. In every run
this difference  changes only where a critical point crosses the boundary (Supplementary Note 3).
The rings split and merge along $z$, so we follow them on the flux ridge, the curve along which $\Psi$ takes its
largest value at each height $z$. The value of $\Psi$ along the ridge defines the ridge profile $\max_r\Psi(r, z)$.
The magnetic axes are its maxima and the saddles between them are minima. The flux surfaces inside the
separatrix through a saddle form a magnetic island, whose depth is the flux difference between the axis and the
saddle. We count an O--X pair as resolved when its island is deeper than $10^{-3}\Psi_{\max}$, with $\Psi_{\max}$ the
peak flux, and wider than a few grid cells (Methods).

Figures~\ref{fig1}c,d and \ref{fig2}a,b,e,f show how the torus splits before the release.
The magnetic field delays the split and changes its kind. At $\eta = 0.05$ the split time rises steadily
with the coupling. At $\eta = 0.01$ it jumps between $\eB = 50$ and 100 and then approaches the release,
which the torus at $\eB = 1000$ reaches unsplit (Fig.~\ref{fig1}c). The delay comes from magnetic tension, and
it is shorter at $\eta = 0.05$, where more of the field has diffused away.

At $\eta = 0.05$ and $\eB = 1$ the original magnetic axis, which sits close to the stagnation plane of the strain,
is displaced to one side as a saddle and a second, nearly equal axis appear on the other (Fig.~\ref{fig2}a and
Supplementary Movie 1). The island depth grows approximately like $(t - t_b)^2$ and the axis--saddle separation like
$(t - t_b)^{1/2}$ (Fig.~\ref{fig2}e). This is a pitchfork-like split, unfolded only slightly, and the
daughters become equal to within a percent as they grow. At $\eB = 50$ the pinch has moved to one side of the original
axis. A saddle and a new axis are born together off the stagnation plane (Fig.~\ref{fig2}b), the depth grows
approximately like $(t - t_b)^{3/2}$ (Fig.~\ref{fig2}f), and the new island is born much shallower than the original
one. The ratio of the two island depths just after the birth, near 1 for the pitchfork and near 0 for the fold, follows this
change across the whole scan. At this diffusivity it falls from the pitchfork band into the fold band as the coupling
grows, and stays there (Fig.~\ref{fig1}d).

The pitchfork-like split is the kinematic limit, and it becomes an exact pitchfork when the offset and the swirl are
removed (Methods). In contrast, the fold arises from the response of the field, not from an asymmetry of the imposed flow.
The swirl winds up a toroidal field that is parallel to the twist of the ring on one side of the midplane and
opposed to it on the other. The magnetic pressure of the toroidal field is higher on the side where
the two add. The pinch occurs on the other side, where the twist is partly unwound. Reversing the
twist of the seed mirrors the split, and removing the offset of the stagnation plane leaves it essentially
unchanged (Methods).

At $\eta = 0.01$ the ratio follows the same descent to $\eB = 50$ and then turns back towards central splitting
(Fig.~\ref{fig1}d). These late splits, at $\eB = 100$--500, happen at the very end of the collapse. At $\eB = 200$--500 
the torus is compressed into a thin, nearly symmetric sheet about the midplane, and the pinch is central once more. Their timing
and symmetry are resolution-sensitive (open circles in Fig.~\ref{fig1}c,d and Methods). Central splits therefore
occur in the kinematic limit and in the late pinch of a strongly compressed sheet, with the fold in between. At
$\eB = 1000$ and $\eta = 0.01$ the compressed torus develops lobes but never pinches through.
By the release the field has lost much of its magnetic helicity, retaining 9--13\% of the initial value
at $\eta = 0.05$ and 44--64\% at $\eta = 0.01$.

\subsection*{A cusp organizes the splits}

\begin{figure*}[tbp]
\centering
\includefigure[\textwidth]{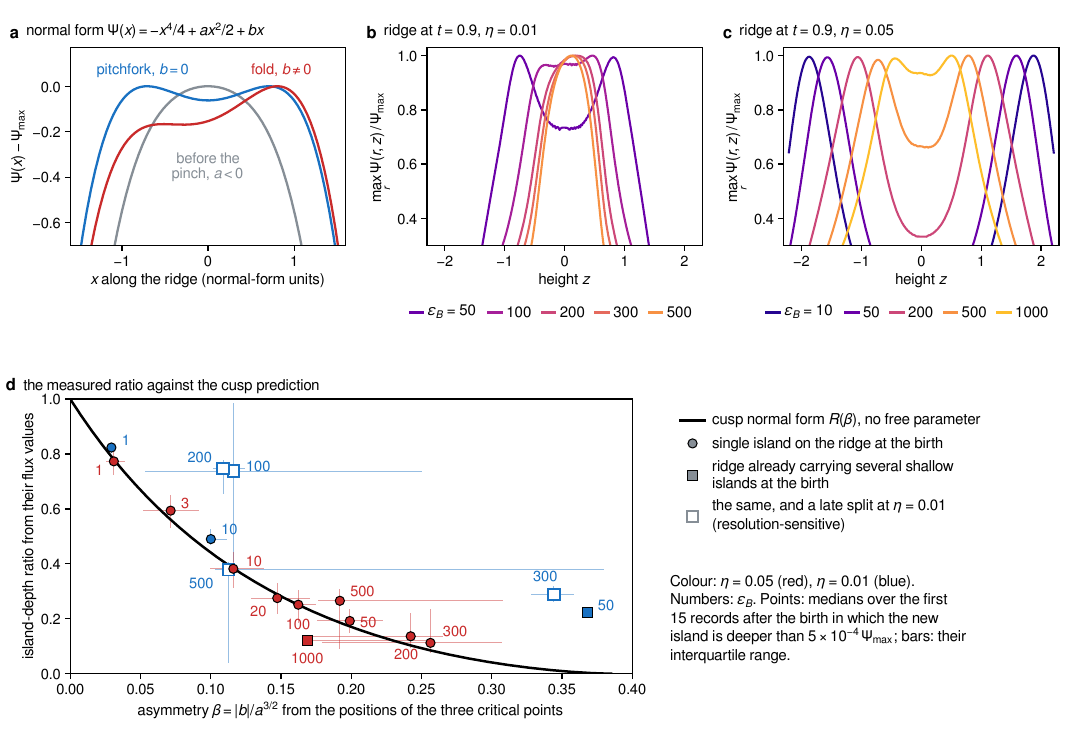}
\caption{\textbf{The cusp description and its test for single-island splits.}
\textbf{Panel a}, the flux along the ridge near the pinch in the normal form $-x^4/4 + ax^2/2 + bx$, before the pinch
($a < 0$, grey), at the symmetric pinch ($b = 0$, blue), a pitchfork, and at the asymmetric pinch ($b\ne0$, red), a
fold.
\textbf{Panel b}, the measured ridge profile at $t = 0.9$ for $\eta = 0.01$. At $\eB = 50$ the torus split at
$t = 0.70$. At $\eB = 200$--500 it has been compressed into a flat sheet, symmetric to within 1--2\% about its centre,
that pinches centrally at $t_s\approx0.96$--1.0.
\textbf{Panel c}, the same for $\eta = 0.05$. The splits are earlier and one-sided, and no flat sheet forms, except at
$\eB = 1000$, where the top is broad and slightly hollow when it pinches.
\textbf{Panel d}, the island-depth ratio, from the flux values of the two axes and the saddle, against the asymmetry
$\beta$, from their positions. The line is the cusp prediction $R(\beta)$. Points are medians over the first 15
records after the birth in which the new island is deeper than $5\times10^{-4}\Psi_{\max}$, and bars are their
interquartile ranges. Circles are births of a single island on the ridge, all the splits at $\eta = 0.05$ up to
$\eB = 500$ and at $\eB = 1$ and 10 for $\eta = 0.01$. Along their sequence at $\eta = 0.05$ the birth moves from
$t = 0.31$ to 0.78 and $\beta$ ranges from 0.03 to 0.26, rising with the coupling though not monotonically. 
Squares are births on a ridge that already carries several
shallow islands, filled for $\eB = 50$ at $\eta = 0.01$ and $\eB = 1000$ at $\eta = 0.05$, and open for the late,
resolution-sensitive splits at $\eta = 0.01$ ($\eB = 100$--500). Four of them lie 0.2--0.4 above the curve, the one
at $\eB = 1000$ lies 0.1 below it, and the one at $\eB = 500$, $\eta = 0.01$ falls on it with the widest spread.
Red, $\eta = 0.05$. Blue, $\eta = 0.01$. Numbers, $\eB$.}
\label{fig3}
\end{figure*}

The pitchfork and the fold are not independent, and a single local description governs both (Fig.~\ref{fig3}). Near
the pinch the ridge profile $\max_r\Psi$ passes from one maximum to two through a degenerate critical point at which
its curvature vanishes. After a shift and a rescaling of height and flux, the leading local form in the height $x$
along the ridge is $\Psi(x) = -x^4/4 + ax^2/2 + bx$ (Fig.~\ref{fig3}a), the two-parameter family of the cusp
catastrophe\cite{thom1975}. The coefficient $a$ is the curvature of the top at its centre $x = 0$. The compression drives it from negative
values, a rounded maximum, through zero to positive values, where two maxima become possible. The coefficient $b$
tilts the top and measures its asymmetry, and it decides which of the two splits occurs. With $b = 0$ the top is flat
when $a = 0$, the central maximum turns into a saddle as soon as $a$ becomes positive, and two equal maxima appear on
either side. This is the pitchfork, the symmetric case of the cusp. With $b\ne0$ the original maximum persists, and a saddle and a
new maximum are born together on one side once $a^3 = \tfrac{27}{4}b^2$. This is the fold, the generic case.

The magnetic field supplies $b$. In the kinematic limit $\mathbf B$ is carried passively by the imposed flow, which
is symmetric about the stagnation plane up to the offset $j = 0.001$. Thus $b$ is small and the split is close to a
pitchfork. As the coupling grows, the Lorentz force drives a response that opposes the pinch. This response is not
symmetric about the midplane, because the toroidal field wound by the swirl reinforces the twist of the ring on one
side and opposes it on the other. This up--down antisymmetric flow grows with the coupling and increases $|b|$, so the
depth ratio of the new islands falls from near 1 towards 0 as the split turns into a fold.

We assume that the quartic holds in the measured height coordinate, and we check this by comparing the positions of
the critical points with the island depths and by fitting the profile directly. The three critical points of the ridge
just after the birth, the two magnetic axes and the saddle between them, are then the roots of $x^3 - ax - b = 0$ up
to a shift and a scale. Their positions fix the asymmetry $\beta = |b|/a^{3/2}$, which does not depend on the scale.
Their flux values give, independently, the ratio of the smaller to the larger island depth. Because the quartic can be
rescaled to $a = 1$, the ratio it predicts depends on $\beta$ alone. We define $R(\beta)$ as the ratio of the smaller
to the larger island depth of $-x^4/4 + x^2/2 + \beta x$. It has no free parameter. It falls from $R(0) = 1$ at the
pitchfork to $R\to0$ at the fold, reached at $\beta = (4/27)^{1/2}$, where the new island has zero 
depth (Supplementary Note 6). Figure~\ref{fig3}d compares the measured ratio with $R(\beta)$ for every split. The measured
ratio and $\beta$ are medians over a short window after the birth (Methods), so that the splits are compared at the
same stage of the pinch although the births are spread over half the collapse. Every split that is born as a single
island on the ridge (circles) lies within 0.08 of $R(\beta)$, with the coupling and the diffusivity entering only
through $\beta$. Along the sequence at $\eta = 0.05$ the measured
asymmetry rises with the coupling, though not monotonically, and the points move down the curve from the pitchfork
towards the fold. Time-resolved fits of the quartic to the ridge profile in two reruns with dense snapshots
support this picture (Methods and Supplementary Fig.~6).

The description fails where the top of the ridge has flattened and broken into ripples before the first island is
resolved. At $\eta = 0.01$ the field resists the compression long enough for the top to flatten before it pinches.
At $\eB = 200$--500 the pinch is then central again (Fig.~\ref{fig3}b), but the broad flat top develops several
shallow ripples instead of the two maxima of a single quartic, as the time-resolved fit at $\eB = 200$ shows
(Methods). The same happens at $\eB = 1000$, $\eta = 0.05$ (Fig.~\ref{fig3}c). The six births on a rippled ridge
(squares in Fig.~\ref{fig3}d) mostly lie well off the curve.

%
%
%
%
%
%
%
%
%
%
%
%

\subsection*{Four kinds of events}

O--X pair events of three kinds change the structure of the flux ridge, and folds of small amplitude make a fourth
pattern (Fig.~\ref{fig2}).

\emph{Fold.} A flux ridge is pinched by converging flow on one side of an axis. A saddle and a new axis are born together
at $t_b$ and then separate as $(t - t_b)^{1/2}$, while the island deepens as $(t - t_b)^{3/2}$ (Fig.~\ref{fig2}b,f).
This is how the torus splits at intermediate and strong coupling. After the release at $\eB = 1000$ it is also how each
daughter sheds an outer ring at a stagnation point of the rebounding flow (Fig.~\ref{fig2}c).

\emph{Pitchfork.} In the symmetric limit of the same pinch, the original axis becomes a saddle and two equal axes appear
on either side, and the islands deepen as $(t - t_b)^2$. The split at weak coupling is close to this limit
(Fig.~\ref{fig2}a,e).  At $\eB = 1000$, $\eta = 0.05$ the split also ends with one axis on each side, but in another
way (Supplementary Movie 2 shows the three-dimensional run). In the axisymmetric run two folds create new axes above
and below the original one, at $t = 0.862$ and 0.869. At $t = 0.880$ the original axis annihilates with the saddle
below it. The split time of Fig.~\ref{fig1}c, 0.88, is the birth of the island left between the two new axes (Methods).

\emph{Inverse fold.} In the fold reversed in time, two axes are driven together by converging flow. The saddle between
them rises as flux reconnects across it until it meets the weaker axis, and both vanish at $t_a$. The island depth and
the separation fall approximately as $(t_a - t)^{3/2}$ and $(t_a - t)^{1/2}$ (Fig.~\ref{fig2}g). This is how the inner
cores of the daughters merge into the equatorial ring of the triplet after the release at $\eB = 1000$
(Fig.~\ref{fig2}c), and how transient rings disappear elsewhere.

\emph{Shallow islands.} Folds of small amplitude create weak local maxima on the necks and lobes of stretched rings.
These shallow islands reach depths of $10^{-3}$ to a few $10^{-2}$ of $\Psi_{\max}$ and mostly vanish again
(Fig.~\ref{fig2}h). At $\eB = 1000$, $\eta = 0.05$ such islands appear on the neck between the daughters around the
release and on their lobes at $1.25 < t < 1.41$ (Fig.~\ref{fig2}c). At $\eta = 0.01$ and $10\le\eB\le300$ the neck
tears into a chain of several islands that persists to the end of the run (Figs.~\ref{fig2}d,h and \ref{fig4}d, and
Supplementary Movie 3 from the three-dimensional run). Whether this is the plasmoid
instability\cite{loureiro2007} cannot be decided at Lundquist numbers of $10^3$ and one cell per resistive layer.

\subsection*{Regimes after the release}

\begin{figure*}[tbp]
\centering
\includefigure[\textwidth]{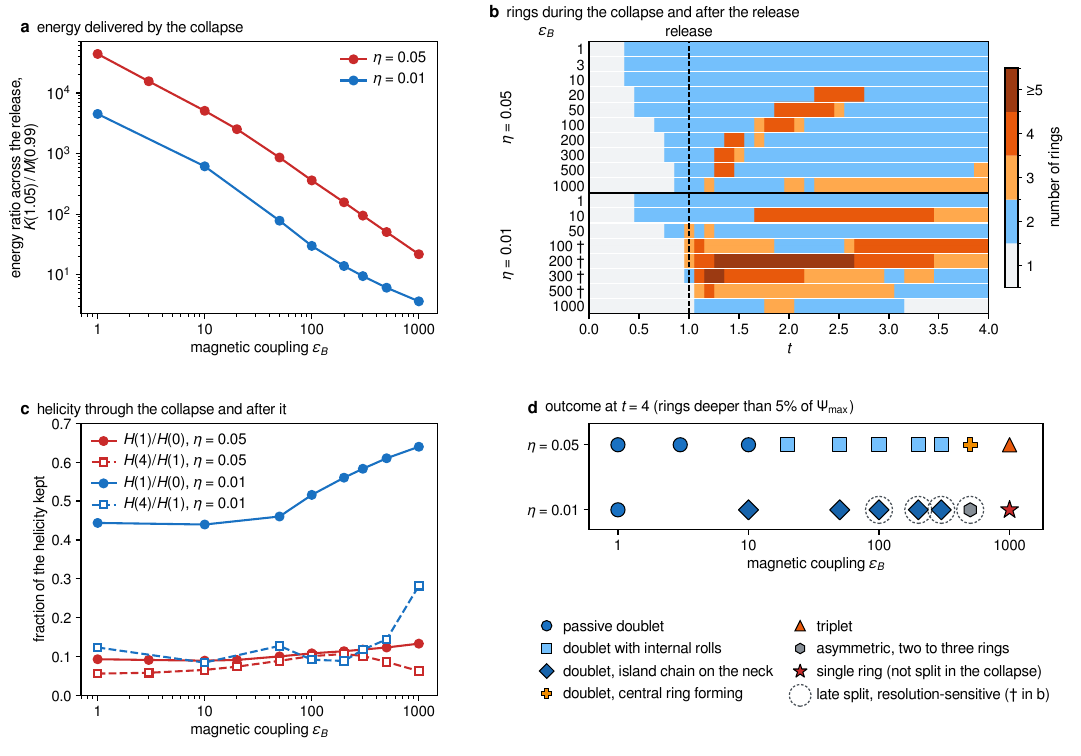}
\caption{\textbf{Regimes after the release.}
\textbf{Panel a}, the energy ratio across the release, the kinetic energy $K$ of the released flow at $t = 1.05$
divided by the magnetic energy $M$ just before the release, at $t = 0.99$, against the magnetic coupling $\eB$. It
spans four decades, from $4\times10^4$ at $\eB = 1$, $\eta = 0.05$ to 4 at $\eB = 1000$, $\eta = 0.01$.
\textbf{Panel b}, number of rings against time for all eighteen runs, counting the main ring and every secondary flux
maximum deeper than 5\% of $\Psi_{\max}$. The triplet forms at $t\approx2.3$ for $\eB = 1000$, $\eta = 0.05$, and the
central ring begins to form at $t\approx3.9$ for $\eB = 500$. Daggers mark the late splits at $\eta = 0.01$, whose
timing and symmetry are resolution-sensitive (Methods).
\textbf{Panel c}, helicity kept through the collapse, $H(1)/H(0)$ (filled circles), and from the release to $t = 4$,
$H(4)/H(1)$ (open squares). After the release the runs keep 6--11\% of the helicity at $\eta = 0.05$ and 8--28\% at
$\eta = 0.01$. The single ring at $\eB = 1000$, $\eta = 0.01$ keeps 28\%, twice as much as any other run.
\textbf{Panel d}, the outcome at $t = 4$ in the $(\eB, \eta)$ plane (Supplementary Figs.~1--3). Passive doublet ($\eB\le10$ at $\eta = 0.05$,
$\eB = 1$ at $\eta = 0.01$), doublet with internal rolls ($20\le\eB\le300$, $\eta = 0.05$), doublet with an island
chain on the neck ($10\le\eB\le300$, $\eta = 0.01$), doublet with a central ring forming ($\eB = 500$,
$\eta = 0.05$), triplet ($\eB = 1000$, $\eta = 0.05$), asymmetric two-to-three-ring evolution ($\eB = 500$,
$\eta = 0.01$) and the surviving single ring ($\eB = 1000$, $\eta = 0.01$). Dashed circles mark the late,
resolution-sensitive splits (daggers in Panel b).}
\label{fig4}
\end{figure*}

After the release at $t = 1$ the state left by the collapse evolves freely (Methods). What
follows depends on the protocol through the kinetic energy of the released flow. It
also depends on the viscosity, which at $\nu = 0.1$ keeps the released flow laminar. A much lower viscosity could
make the flow turbulent and change the weak-field outcomes, which we have not tested.
Figure~\ref{fig4} shows what becomes of the rings.

We define the energy ratio across the release as the kinetic energy of the released flow at $t = 1.05$ divided by
the magnetic energy at $t = 0.99$, just before the release. It spans 
four decades across the scan (Fig.~\ref{fig4}a). Together with the magnetic
diffusivity, it orders the runs into a sequence of regimes (Fig.~\ref{fig4}b,d). At weak coupling the field is passive.
The two daughters, nearly equal by then, are carried by the axial jets to $|z|\approx4$--5 and remain a doublet.
The swirl winds their toroidal field against its original twist, and partly reverses it. At $\eB = 1$, $\eta = 0.01$ a
quarter of $|B_\phi|$ is reversed by $t = 4$. At intermediate coupling and $\eta = 0.05$ the daughters develop
internal rolls that briefly register as extra rings before relaxing to a clean doublet. At $\eta = 0.01$ the
doublet is instead accompanied by a chain of islands on the neck between the daughters (Fig.~\ref{fig2}d,h).

At the strongest coupling and $\eta = 0.05$ the rebounding flow reverses the collapse. Fluid moves in along the axis
towards the origin and out along the midplane, with three stagnation points on the axis. Each daughter lies across an
outer stagnation point, where the axial flow diverges, and sheds an outer ring in a fold. The inner cores are swept to
the origin and merge in an inverse fold (Fig.~\ref{fig2}c,g). The result is a triplet of two compact tori drifting
apart with a flat equatorial ring between them (Fig.~\ref{fig5}b and Supplementary Movie 2, from the
three-dimensional run). The triplet is complete at $\eB = 1000$, but at $\eB = 500$ the central ring has only started
to form when the run ends at $t = 4$. At $\eB = 1000$ and $\eta = 0.01$, with the smallest energy ratio, the torus has
not split before the release, and the released flow displaces the whole ring along the axis. The ring sheds weak
remnants and survives to $t = 4$ as a single ring, keeping more of its helicity than any other run
(Fig.~\ref{fig4}c,d).

Lowering the magnetic diffusivity thus preserves more of the field. At the strongest coupling the forming central ring
and the triplet of $\eta = 0.05$ give way to an asymmetric evolution with two to three rings at $\eB = 500$ and to the
single ring at $\eB = 1000$. At intermediate coupling the thin neck between the daughters tears, and island chains
replace the internal rolls of the $\eta = 0.05$ doublets. After the release the helicity decays resistively (Methods),
so the regimes describe decaying fields (Fig.~\ref{fig4}c). The $\eta = 0.01$ runs keep enough field for the
distinctions to be meaningful to $t\approx3$.

%
%
%
%
%
%
%
%
%
%
%

\subsection*{Three dimensions}

\begin{figure*}[tbp]
\centering
\includefigure[\textwidth]{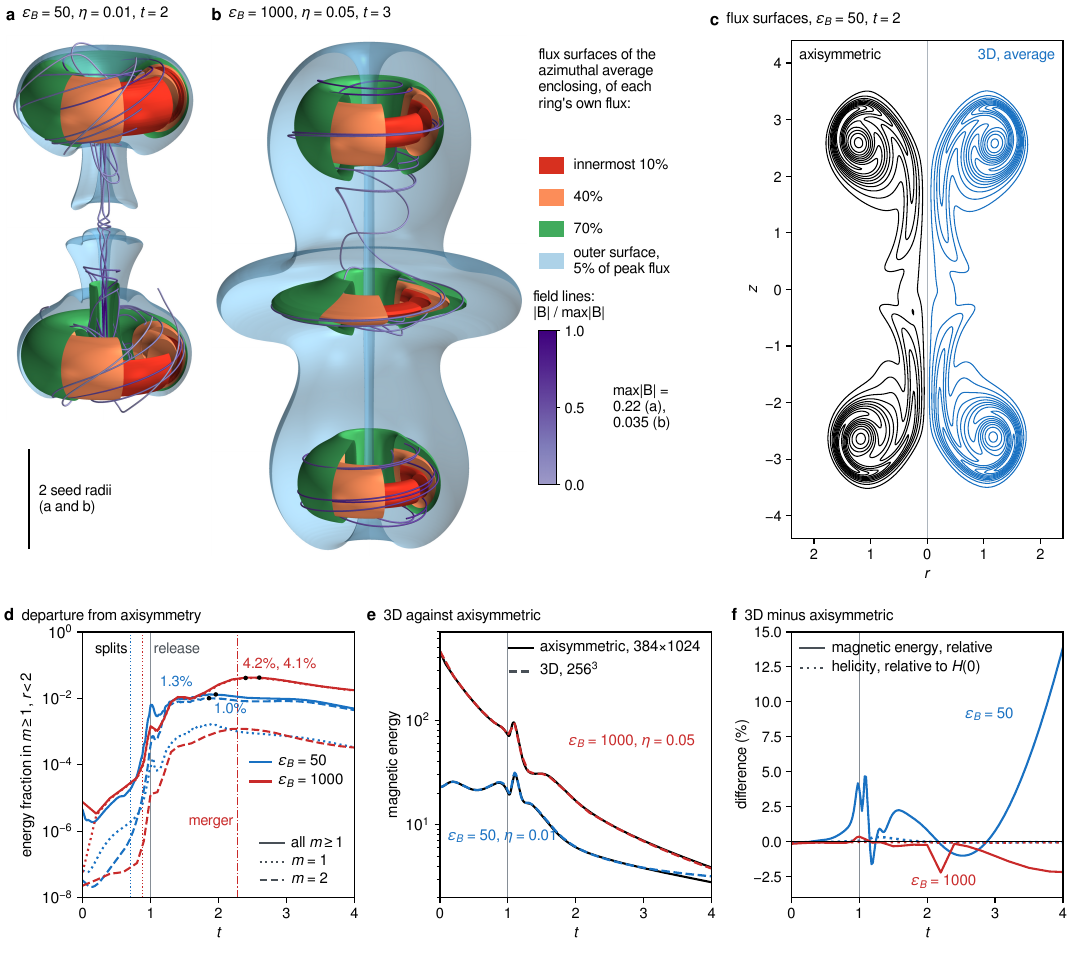}
\caption{\textbf{Three dimensions.} Two cases of the scan repeated in a $256^3$ Cartesian simulation with a random
solenoidal magnetic perturbation of relative amplitude $10^{-3}$ added to the seed.
\textbf{Panel a}, $\eB = 50$, $\eta = 0.01$ at $t = 2$: the two daughters of the fold and the neck between them.
\textbf{Panel b}, $\eB = 1000$, $\eta = 0.05$ at $t = 3$: the triplet, two compact tori and the flat equatorial ring formed by
the merger of the daughters' cores.
In Panels a and b, drawn to the same scale, the surfaces are flux surfaces of the azimuthally averaged field
in the colour code of Fig.~\ref{fig1}a, applied to each ring's own flux, revolved about the axis and cut open. The field
lines (purple, shaded by $|\mathbf B|$ relative to its maximum in the snapshot) are integrated in the full
three-dimensional field from the 40 and 70\% surfaces of each ring and, in Panel a, from the neck between the
daughters, where they braid around one another.
\textbf{Panel c}, flux surfaces at $t = 2$ for $\eB = 50$: the axisymmetric run (black, left) and the azimuthal average of the
three-dimensional field (blue, right).
\textbf{Panel d}, fraction of $|\mathbf B|^2$ sampled inside $r < 2$ and $|z| < 0.8L_z$, with $L_z$ the half-height of the
box (Methods), in all azimuthal modes $m\ge1$ (solid), in $m = 1$ (dotted)
and in $m = 2$ (dashed), for both runs. The seeded $10^{-6}$ is stretched by the collapse to $10^{-3}$--$10^{-2}$ at the
release, grows further after it, reaches at most 1.3\% (dominated by $m = 2$ while the neck tears, $\eB = 50$) and 4.2\%
(dominated by $m = 1$ during the merger of the cores, $\eB = 1000$), and decays thereafter. Vertical lines: the splits,
the release and the merger.
\textbf{Panel e}, magnetic energy of the three-dimensional runs (dashed) against the axisymmetric ones (black).
\textbf{Panel f}, their relative difference (solid), and the difference of the helicities relative to the initial helicity
(dotted).}
\label{fig5}
\end{figure*}

The bifurcations and regimes described so far come from the axisymmetric simulations. A compressed torus can break
that symmetry by tilting or kinking, and this is to be expected above all during the reconnections that split and
merge it. We therefore repeated two cases in three dimensions (Methods), the fold with the island chain at
$\eB = 50$, $\eta = 0.01$ and the triplet at $\eB = 1000$, $\eta = 0.05$ (Fig.~\ref{fig5}a,b and Supplementary
Movies 3 and 2). Both follow the axisymmetric runs through $t = 4$ (Fig.~\ref{fig5}c--f). The helicity agrees to
within 0.4\% of its initial value throughout. The magnetic energy agrees to within 5\% up to $t = 3$, and at
$\eB = 50$ the difference grows to 14\% by $t = 4$ (Fig.~\ref{fig5}e,f). The flux surfaces of the azimuthal average
lie on those of the axisymmetric solution, down to the neck between the daughters (Fig.~\ref{fig5}c).

The seeded perturbation is stretched by the collapse, continues to grow after the release, and  peaks at
the main reconnection event of each case before it decays (Fig.~\ref{fig5}d). At $\eB = 50$ the $m = 2$ mode dominates
as the neck between the daughters tears, and at $\eB = 1000$ the $m = 1$ mode dominates during the merger of the inner
cores into the central ring. The other modes stay below 0.2\%, except $m = 4$, to which the Cartesian lattice also
contributes and which peaks at 0.4\% at the release at $\eB = 50$. Continued to $t = 8$ in a larger box, the
$\eB = 50$ run decays with $M/|H|$ falling from 300 to 240, as in selective decay, and its non-axisymmetric fraction falls
tenfold (Supplementary Note 7 and Supplementary Movie 3). In both cases the departure from axisymmetry thus stays
below 5\% of the magnetic energy inside $r < 2$ and $|z| < 0.8L_z$ (Fig.~\ref{fig5}d). The large-scale splits and mergers
persist, at least for the perturbation and lattice tested, which does not resolve the reconnection layers (Methods).
%
%
%
%
%
%
%
%
%
%
%
%

\section*{Discussion}

A magnetic flux ring driven by the collapse profile of a reported Navier--Stokes singularity does not simply
compress. It comes apart through bifurcations of its flux function, and the magnetic feedback selects which one. For
splits born as a single island on the ridge this selection is quantitative, since the positions of the critical points
predict the island-depth ratio with no free parameter.

The local pitchfork and fold geometries do not require a singular strain, and the cusp describes the pinch of any
rounded flux ridge whose profile admits a local quartic normal form. Similar pitchfork-like splits at weak coupling
and folds at stronger coupling may therefore occur in other converging, rotating flows, whenever the rotation winds
the toroidal field of the ring unequally on the two sides of the stagnation plane. Whether such flows reproduce the
full sequence found here remains to be tested.

 What the collapse profile contributes is the time dependence of the compression, through the $1/(1-t)$ growth of its
unregularized strain. Magnetic tension delays the strong-field splits, and the continued compression flattens the
torus into a thin sheet, which at $\eta = 0.01$ pinches centrally near $t = 1$ for $\eB = 200$--500. The
regularization lowers the midplane strain of the imposed flow by 0.8\% at $t = 0.6$ and by 2.5\% at $t = 0.78$. All
splits at $\eta = 0.05$ up to $\eB = 500$ are born by then. They should therefore depend less on $\tau_f$ than the
later splits, which occur where the regularization is strong. Neither dependence has been measured. How the absence of a split at
$\eB = 1000$, $\eta = 0.01$ depends on $\tau_f$, and whether magnetic tension can prevent the split in the
unregularized limit, are questions for a smaller $\tau_f$ and finer lattices.

The spatial cutoff confines the meridional flow of the imposed
profile, while the swirl remains unconfined and its amplitude is a model choice. The forced magnetohydrodynamic
equations we solve do not require the admissible-stress condition of the full construction, and the shear of our
swirl does not satisfy it. The evolution for $t > 1$ depends in addition on the release, and a bounce remains to be
explored. Near $t = 1$ the thinnest current sheets of the $\eta = 0.01$ runs approach the lattice scale, so the number
of shallow islands and the timing and asymmetry of the late flat-sheet splits remain uncertain (Methods). The
three-dimensional runs are two to three and a half times coarser than the axisymmetric ones, with cells as wide as
the resistive layer at $\eta = 0.05$ and wider at $\eta = 0.01$.

Compact-torus compression experiments\cite{brown2015,lindemuth1983,kaur2018} could test the diffusivity-dependent
delay of the split, the transition from central to one-sided pinching of a rounded torus, the local cusp relation, and
the formation of a triplet after the release at strong coupling. For illustration, a deuterium spheromak of radius
10\,cm compressed over 10\,$\mu$s, with a field of 0.5\,T and an ion density of $10^{21}$\,m$^{-3}$, has
$\eB\approx600$ in our units. At an electron temperature of 20\,eV a parallel Spitzer estimate with a Coulomb
logarithm of 10 gives $\eta\approx0.005$, half the lowest value of our scan, and the scalar viscosity remains a model
choice. Compressibility, anisotropic transport and the kinetic effects that our model omits would first have to be
assessed for the device at hand. In the solar corona, photospheric motions stress the structures traced by coronal
loops, and reconnection transfers flux between them\cite{reale2014,priest2000}. There the field dominates the flow and
the driving is slow compared with the Alfv\'en time, so only the strong-coupling end of our scan has a counterpart.
The coronal structures are also three-dimensional, anchored in the photosphere, and at Lundquist numbers many orders
of magnitude larger than ours. Whether the competition between converging flows and magnetic tension and the local
cusp description remain useful there is a question that observations and coronal simulations could test.
\section*{Methods}

\paragraph{Model and numerics.} In axial symmetry, equations (\ref{eq:ns}) and (\ref{eq:ind}) are solved for the
response flow $\mathbf v$ and the field $\mathbf B$ in the background of the prescribed flow $\mathbf U(r, z, t)$. The
pressure $p_U$ of the imposed flow in the force $\mathbf f$, $\nabla^2 p_U = -\nabla\cdot[(\mathbf U\cdot\nabla)\mathbf U]$,
makes $\mathbf f$ divergence-free. The cylindrical box $0 < r < 6$, $|z| < 8$ carries a $384\times1024$ cell-centred
lattice, with perfectly conducting and impermeable walls. The scheme is third-order upwind in advection, second 
order in diffusion and third-order SSP Runge--Kutta in time, with the Alfv\'en speed in the CFL condition 
and an exact Poisson solve for the stream function. Every cutoff, taper and ramp uses the same smooth 
step $C(x)$, which equals 1 for $x\le0$, 0 for $x\ge1$ and $[1 + \exp(-1/x + 1/(1 - x))]^{-1}$ in between, 
with all derivatives continuous. 

\paragraph{Imposed flow.} The meridional flow derives from the Stokes stream function $\psi_U = r^2S_U$, with
$U_r = -r^{-1}\partial_z\psi_U$ and $U_z = r^{-1}\partial_r\psi_U$. Here
$S_U = \tfrac12(4Z + j)/\lambda^{1+\delta}$ is the reduced stream function of the reference branch of the
reconstruction\cite{leiren2026}, with $\delta = 0.005$ and $j = 0.001$. $R = r^2/(2\lambda^2)$ and
$Z = z/\lambda^{1-\delta}$ are the similarity coordinates, and the scale $\lambda(t, z)$ of the core is defined by
$\lambda^2 - \lambda^{2\delta}z^2 = \tau_e$. $\tau_e = \sqrt{(1-t)^2 + \tau_f^2}$ replaces the $1 - t$ of the
unregularized profile and the regularization time $\tau_f = 0.05$ keeps $\lambda$ finite. $S_U$ is multiplied by
the cutoff $C(r^2/(2R_c\tau_e) - 1)$, which equals 1 inside the similarity radius $R_c = 16$ measured with the
midplane scale and 0 beyond $2R_c$. The swirl is
$u_\theta = A f(Z) g(R)/\lambda^{1+\delta}$ with $f = 1/(1+Z^2)$,
$g = [R/(R + X_c)]^{1/2}[(R + X_c)/X_R]^{1/10}[1 + (R/X_R)^2]^{-3/10}$, $X_c = 0.5$, $X_R = 8$ and $A = 2$, 
not subject to the cutoff (Supplementary Note 4).

\paragraph{Seed, units and dimensionless numbers.} Seed is the Taylor spheromak
$\Psi\propto r^2 j_1(\lambda_Ts)/(\lambda_Ts)$. Here $s = \sqrt{r^2 + z^2}$ and  $j_1$ the spherical Bessel function of
order one, $\lambda_T = 4.4934$, truncated smoothly between $s = 0.7$ and 1 by the step $C$ and normalized to unit
peak field. Its toroidal field is $B_\phi = -\lambda_T\Psi/r$, with the sign chosen so that the twist is parallel to
the field wound by the swirl in the upper half-space (Supplementary Note 2). The
Lundquist number $S = \sqrt{\eB}\,B_{\max}/\eta$, with $B_{\max}$ the peak field at $t = 1$, ranges from 5 to 4277
across the scan (Supplementary Note 1 and Supplementary Tables 1 and 2), the Reynolds number of the imposed 
flow at the release is about 500 and the magnetic Prandtl number is
$\nu/\eta$. Control runs at $\eB = 50$, $\eta = 0.05$ without the offset ($j = 0$) and with the twist of the seed
reversed reproduce and mirror the reference split. An exactly symmetric run at $\eB = 1$, without offset and
swirl, splits as an exact pitchfork (Supplementary Note 2, Supplementary Table 3 and Supplementary Fig. 4).

\paragraph{Release.} In the core of the collapse the swirl and axial velocities of the profile scale as
$\lambda^{-(1+\delta)}$ and the volume as $\lambda^{3-\delta}$, with $\lambda^2\propto1 - t$. The kinetic energy
inside the core vanishes like $(1-t)^{(1-3\delta)/2}$ for $\delta < 1/3$ while the velocity and its gradients
diverge. The regularized imposed flow at $t = 1$ keeps a kinetic energy of $2.4\times10^3$ at
$\tau_f = 0.05$. The total kinetic energy at the release, with the response, is $2.3$--$2.7\times10^3$ across the
scan (Supplementary Note 4 and Supplementary Table 5).  At $t = 1$ the vorticity and angular velocity of the imposed flow are added to the
response, $\mathbf U$ is set to zero and the magnetic field is not modified. The velocity and the kinetic energy
change by less than one part in $10^{13}$, the level of round-off error, and the acceleration jumps by the
solenoidal part of the force (Supplementary Note 5). A control at $\eB = 1000$, $\eta = 0.05$ with the force 
switched off smoothly between $t = 1$ and 1.05 keeps the kinds and order of the pair events after the release. 
The events occur later, by up to 0.4 collapse times for the merger of the cores, at $t = 2.65$--2.70 instead of 
2.28, and the triplet forms in both runs (Supplementary Note 5).

\paragraph{Critical points and islands.} The critical points of the flux ridge $\max_r\Psi(r, z)$ are tracked every
two time steps. Up to the release the radial maximum is unique and nondegenerate, with
$-r_*^2\partial_r^2\Psi > 3.7\Psi$ at the ridge radius $r_*(z)$, except in the compressed sheets at $\eta = 0.01$ at
$t = 1$. The profile $F(z) = \Psi(r_*(z), z)$ is therefore a single smooth branch, with
$F'' = \Psi_{zz} - \Psi_{rz}^2/\Psi_{rr}$ on the ridge. Its maxima are O points of $\Psi$, its minima are X points,
and its degenerate critical points are the bifurcations of the text. Resolved islands are deeper 
than $10^{-3}\Psi_{\max}$ and wider than three cells (Supplementary Note 3).  The birth
time $t_b$ of Fig.~\ref{fig2}e,f is where a straight line fitted to the squared axis--saddle separation over the
first 25 resolved records reaches zero. This fit presumes the square-root law. The annihilation time $t_a$ of
Fig.~\ref{fig2}g is found in the same way from the last 25 records. The exponents are the slopes of the depth and of
the separation on logarithmic axes over $0 < t - t_b < 0.05$.
Joint fits of $t_b$ and the exponent change the depth exponents by up to 0.48, so the exponents 
are finite-window estimates consistent
with the normal form rather than its asymptotic values (Supplementary Note 6 and Supplementary Table 7).
The helicity $H = 4\pi\!\int\Psi B_\phi\,dr\,dz$ is gauge invariant, since no field crosses the walls, and after the
release it decays only resistively (Supplementary Note 5).

\paragraph{Split statistics and the cusp test.} The split time $t_s$ is the first detection, at the noise floor of
the ridge, of the first island that grows deeper than $3\times10^{-3}\Psi_{\max}$, and it differs from $t_b$ by up to
0.005. We follow this island back along the tracks, as the deepest island at nearly the same height, to its first
appearance. At $\eB = 1000$, $\eta = 0.05$ the islands of the two folds stay within the ripples
of the top, so $t_s = 0.88$ marks the island between the new axes. The cusp test (Fig.~\ref{fig3}d) uses the medians of
the ratio and of $\beta$ over the first 15 records in which the island is deeper than $5\times10^{-4}\Psi_{\max}$.
The threshold is half the resolved one because the quartic holds only close to the birth, and with $10^{-3}\Psi_{\max}$ instead
the single-island births lie within 0.11 of $R(\beta)$ rather than 0.08 (Supplementary Note 6 and Supplementary
Table 6). A birth counts as single-island when no other secondary island deeper than $2\times10^{-4}\Psi_{\max}$ is on the
ridge at the start of this window. As an independent check, the quartic
was fitted to the ridge profile of two reruns with snapshots every 0.005 time units through the split
(Supplementary Fig.~6). For the fold at $\eB = 100$, $\eta = 0.05$ the fitted coefficients meet the fold condition at
$t = 0.546$, against 0.545 and 0.548 for the first detection and the fitted birth in the tracks; the predicted
island depth runs ahead of the measured one, by a factor of two at $t = 0.56$. For the late split at $\eB = 200$,
$\eta = 0.01$ the top flattens at $t = 0.917$ and breaks into several shallow maxima before the first resolved
island appears at $t = 0.959$.

\paragraph{Numerical convergence.} On $128\times256$ and $256\times512$ lattices of a smaller box the split time at
$\eB = 10$, $\eta = 0.05$ changes by 0.001 and the island-depth exponents by less than 0.05. The resistive 
layer $\sqrt{\eta\tau_f}$ at $t = 1$ spans about three cells at $\eta = 0.05$ and about one
cell at $\eta = 0.01$.  For three runs the energy and helicity budgets were integrated in time with the same discrete
operators (Supplementary Note 5 and Supplementary Fig.~5). The helicity budget closes to within 0.8\% of the helicity dissipated. The magnetic
energy budget closes to within 1.3\% of the resistive dissipation at $\eta = 0.05$ and 4.7\% at $\eta = 0.01$. After
the release, where no imposed flow enters, the total energy budget closes to within 0.5\% of the energy dissipated.
A repeat of the fold at $\eB = 50$, $\eta = 0.05$ on a $512\times1365$ lattice
reproduces the times at which the new island reaches each depth from $10^{-4}$ to $10^{-2}\Psi_{\max}$ within 0.0004,
the saddle heights within 0.005 and the island-depth ratio 0.19, while the first detection at the noise floor shifts
by 0.003 (Supplementary Note 3 and Supplementary Table 4). This holds for that fold, not for every event of the 
scan. In a repeat of the $\eB = 100$, $\eta = 0.01$ run on the finer lattice the
energies, the helicity and the ring history agree, but the split comes later and is more central, at $t_s = 0.954$
against 0.921 with an island-depth ratio of 0.50 against 0.29. Small perturbations decide where and when a flat top
pinches, so we report the late splits at $\eta = 0.01$ as exploratory, marked as such in Figs.~\ref{fig1}c,d,
\ref{fig3}d and \ref{fig4}b,d.

\paragraph{Three dimensions.} A pseudo-spectral code solves equations (\ref{eq:ns}) and (\ref{eq:ind}) in a periodic
Cartesian box with $256^3$ points, $2/3$ dealiasing and SSP-RK3 time stepping with the Alfv\'en speed in the CFL
condition. The background and the seed are the axisymmetric ones, tapered
to zero before the box faces. The boxes are $|x|, |y| < 4$, $|z| < 5.5$ for $\eB = 50$, $\eta = 0.01$ and
$|x|, |y| < 5$, $|z| < 7$ for $\eB = 1000$, $\eta = 0.05$. The resistive layer spans
0.5--0.7 cells at $\eta = 0.01$ and 0.9--1.3 cells at $\eta = 0.05$, so the runs resolve the large-scale structure but
not the current sheets. The perturbation is solenoidal Gaussian noise below the wavenumber $6\pi/L_z$, with $L_z$
the half-height of the box, confined to the seed by a Gaussian window and scaled to a relative $L_2$ amplitude of
$10^{-3}$. The fractions of
$|\mathbf B|^2$ in the azimuthal modes are sampled on cylindrical points equally spaced in $r$, $\phi$ and $z$ inside
$r < 2$ and $|z| < 0.8L_z$. This region holds over 99.9\% of the magnetic energy to $t = 3$ and 76\% at $t = 4$ at
$\eB = 50$, and 98\% at $t = 3$ at $\eB = 1000$. Extending it to the whole box
changes the fractions by less than 3\% of their value, and weighting the samples by volume lowers the peak at
$\eB = 1000$ from 3.4\% to 2.4\%.

\section*{Use of AI tools.} The simulation codes, the analysis scripts and the figure scripts were written with
the assistance of large language models (both Anthropic Claude and OpenAI models) under the
author's direction.  Dialogues with ChatGPT and Claude were used  in particular to clarify
aspects of the reported OpenAI Navier--Stokes construction.
All runs were performed by the author on local hardware, and all code, analyses and figures
were checked by the author, who takes full responsibility for them. AI tools were also used for language editing
of the manuscript.

\section*{Data and code availability}

The axisymmetric and the three-dimensional simulation codes in the versions that produced the runs, the
configuration of every run, the analysis and figure scripts, and the data from which every figure and table of this
paper and of its Supplementary Information are drawn (the time series, the critical-point tracks, the ring records
and the summaries of all runs, the control runs, the three-dimensional diagnostics and the derived numbers) are
available in a single repository at \url{https://github.com/niemiantti/Navier_Stokes/} 
under the MIT licence for the code and the CC BY 4.0 licence for the data. The snapshots of the fields
are not stored. They are regenerated by running the stored configurations with the stored code, as described in the
repository. Source data for Figs.~\ref{fig1}--\ref{fig5} are provided with this paper.

\section*{Acknowledgements}

 This research is supported by the Swedish Research Council under Contract No. 2022-04037 
and by COST Action CA23134 (POLYTOPO).

\section*{Author contributions}

A.J.N. conceived the study, performed the simulations and the analysis, and wrote the paper.

\section*{Competing interests}

The author declares no competing interests.

\section*{Additional information}

\textbf{Supplementary information} is available for this paper, with Supplementary Notes 1--7, 
Supplementary Figs.~1--6, Supplementary Tables 1--7 and Supplementary Movies 1--3. \textbf{Correspondence} and requests for 
materials should be addressed to
A.J.N.

\section*{Supplementary Movies}

All three movies have the same layout. The left panel shows the meridional section of the axisymmetric field
(Movie 1) or of the azimuthal average of the three-dimensional field (Movies 2 and 3), mirrored across the axis.
Each magnetic ring is drawn with four flux layers in the colour code of Fig.~\ref{fig1}a. Red, orange and green
are the surfaces enclosing 10, 40 and 70\% of the ring's own flux about its axis, and blue is the outer surface at
5\% of the peak flux. Thin black lines are flux surfaces and the magnetic axes are gold. The streamlines of the
meridional velocity are thin lines, grey in Movies 1 and 2 and light blue in Movie 3, slightly thicker where the
fluid moves faster. The label gives the time and states whether the flow is imposed or released. The right panel
shows the same flux surfaces revolved about the axis with a staircase cut-away, at the same vertical scale as the
section. Frames are interpolated linearly between the stored times and played at 24 frames per second.

\paragraph{Supplementary Movie 1.} The axisymmetric run at $\eB = 1$, $\eta = 0.05$ from $t = 0$ to 4, the run
of Fig.~\ref{fig2}a, in the window $r < 3$, $|z| < 7$ (6.7~s, 161 frames from the snapshots stored every 0.05
time units). The streamlines are those of the total velocity. Before the release this is the imposed flow plus the
response, and after it the released flow. The movie shows the compression of the torus and the pitchfork-like split
at $t = 0.31$ into two nearly equal daughters. It then shows the release at $t = 1$ and the passive doublet, carried
by the axial jets to $|z|\approx4$ while the swirl partly rewinds its toroidal field.

\paragraph{Supplementary Movie 2.} The three-dimensional run at $\eB = 1000$, $\eta = 0.05$ from $t = 0.1$ to 4,
the run of Fig.~\ref{fig5}b, as its azimuthal average, in the window $r < 5$, $|z| < 7$ (6.5~s, 157 frames from
the averages stored every 0.1 time units). The stored averages of this run contain the field only. The streamlines
are therefore drawn from the velocity of its full snapshots, which are 0.5 time units apart. Before the release
the velocity is the imposed flow at the time of the frame plus the response interpolated between the snapshots.
After the release it is the total velocity interpolated between them. The movie shows the compression of the torus
into a broad sheet and the split by two folds, followed by the annihilation of the original axis at
$t = 0.86$--0.88. It then shows the release, the brief compression of the field by the rebounding flow, the shallow
islands on the lobes and the shedding of the outer rings in folds. The inner cores merge into the central ring near
$t = 2.3$ and form the triplet. The events occur at the times given in the text to within the resolution of the
averages.

\paragraph{Supplementary Movie 3.} The three-dimensional run at $\eB = 50$, $\eta = 0.01$ from $t = 0.1$ to 8,
the run of Fig.~\ref{fig5}a, as its azimuthal average, in the window $r < 3$, $|z| < 7$ (13~s, 317 frames from the
averages stored every 0.1 time units). From $t = 4$ to 8 the run continues in a larger box with the same cells
(Supplementary Note 7). The streamlines are those of the azimuthally averaged velocity stored with the field. The
movie shows the compression of the torus by the collapse, the fold at $t = 0.70$ and the release at $t = 1$. It
then shows the tearing of the neck into the chain of islands, the drift of the daughters along the axis and the
decay of the field to $t = 8$.


\newpage

\begin{center}
{\Large\bfseries Supplementary Information}\\[6pt]
\end{center}

\vspace{4pt}
{\it \noindent This Supplementary Information contains Supplementary Notes 1--7, Supplementary Figures 1--6 and
Supplementary Tables 1--7. The three Supplementary Movies are described in a separate file. Supplementary
Note~\ref{sec:runs} collects the parameters and the results of the eighteen runs of the scan. It also shows every run
at the release and at four later times. Supplementary Note~\ref{sec:asym} isolates the source of the asymmetry of
the fold. Supplementary Note~\ref{sec:res} gives the lattice dependence of two splits and the bookkeeping of the
critical points. Supplementary Note~\ref{sec:flow} describes what the imposed flow keeps of the construction, its
cutoff and its regularization. Supplementary Note~\ref{sec:release} describes the release, the energy and helicity
budgets and a control with a smooth switch-off of the force. Supplementary Note~\ref{sec:cusp} gives the scope and
the robustness of the cusp relation. Supplementary Note~\ref{sec:3d} describes a three-dimensional run to $t = 8$.
The seed radius is the unit of length and the collapse time the unit of time. The seed has unit peak field.}

\section{The runs of the scan}
\label{sec:runs}
\begin{figure}[p]
\centering
\includegraphics[height=0.84\textheight,keepaspectratio]{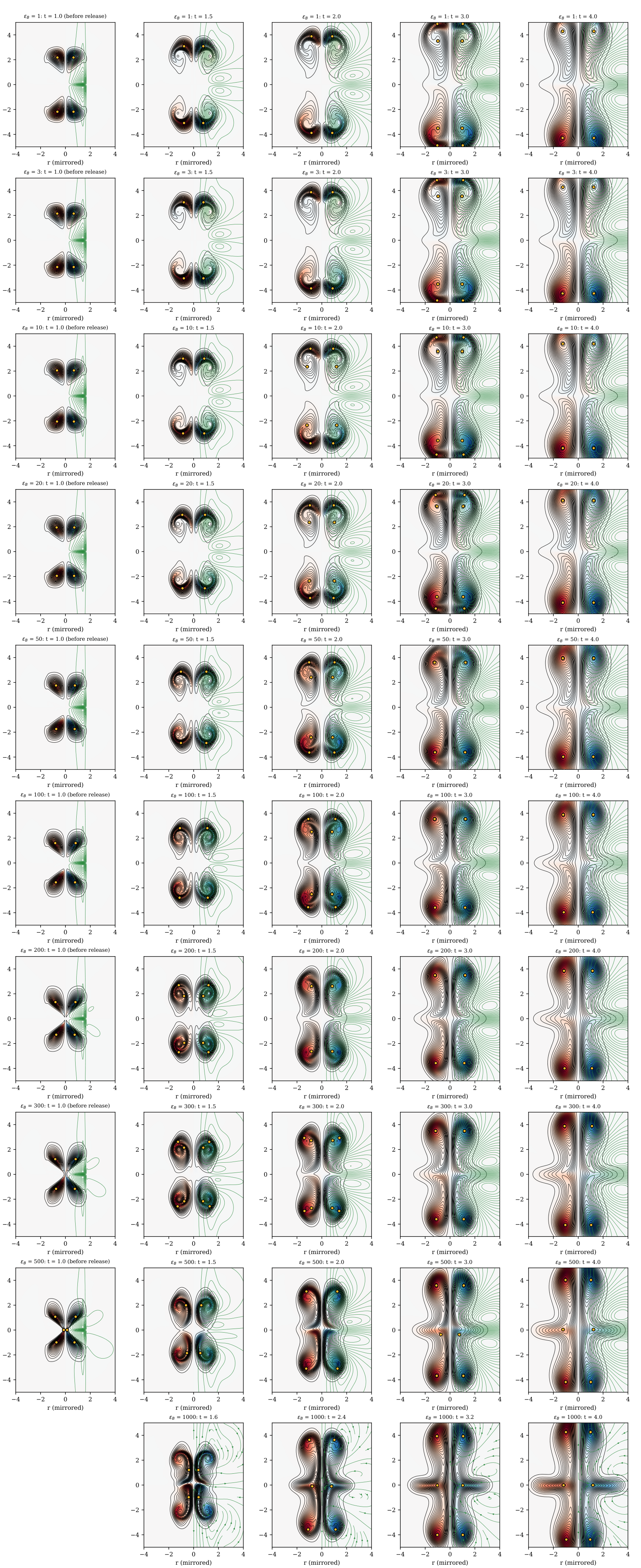}
\caption{\textbf{Snapshots of the $\eta = 0.05$ runs.} Toroidal field $B_\phi$ (colour, red positive and blue
negative), flux surfaces (black), magnetic axes (gold) and the total flow (green streamlines). The times are $t = 1$,
immediately before the release, and $t = 1.5$, 2, 3 and 4. The window is $r < 4$, $|z| < 5$, mirrored across the
axis. The rows are $\eB = 1$, 3, 10, 20, 50, 100, 200, 300, 500 and 1000. The $\eB = 1000$ run was computed in two
parts with a checkpoint at $t = 1.5$. It is shown at $t = 1.6$, 2.4, 3.2 and 4. Up to $\eB = 10$ the daughters
ride the axial jets as a passive doublet. From $\eB = 20$ to 300 they develop internal rolls, which register briefly
as extra rings. At $\eB = 500$ a central ring begins to form late in the run. At $\eB = 1000$ the daughters shed
their outer rings and their cores merge into the triplet.}
\label{fig:snap005}
\end{figure}

\begin{figure}[p]
\centering
\includegraphics[height=0.84\textheight,keepaspectratio]{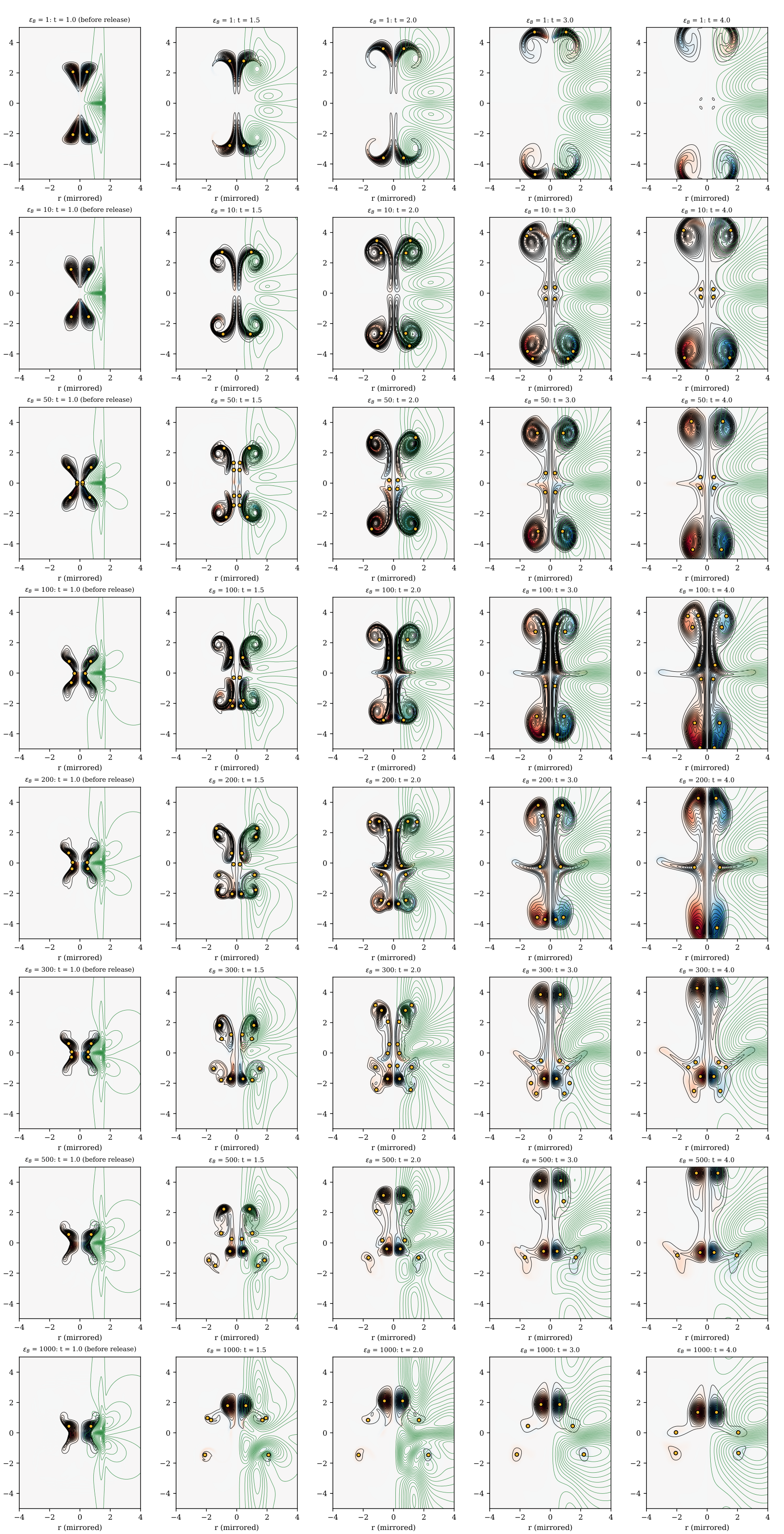}
\caption{\textbf{Snapshots of the $\eta = 0.01$ runs.} As Supplementary Fig.~\ref{fig:snap005}, for $\eB = 1$, 10,
50, 100, 200, 300, 500 and 1000. At $\eB = 1$ the doublet is passive. From $\eB = 10$ to 300 the neck between the
daughters tears into a chain of islands. At $\eB = 500$ the evolution is asymmetric with two to three rings. At
$\eB = 1000$ the torus is never pinched through and survives as a single ring.}
\label{fig:snap001}
\end{figure}

\begin{figure}[p]
\centering
\includegraphics[height=0.8\textheight,keepaspectratio]{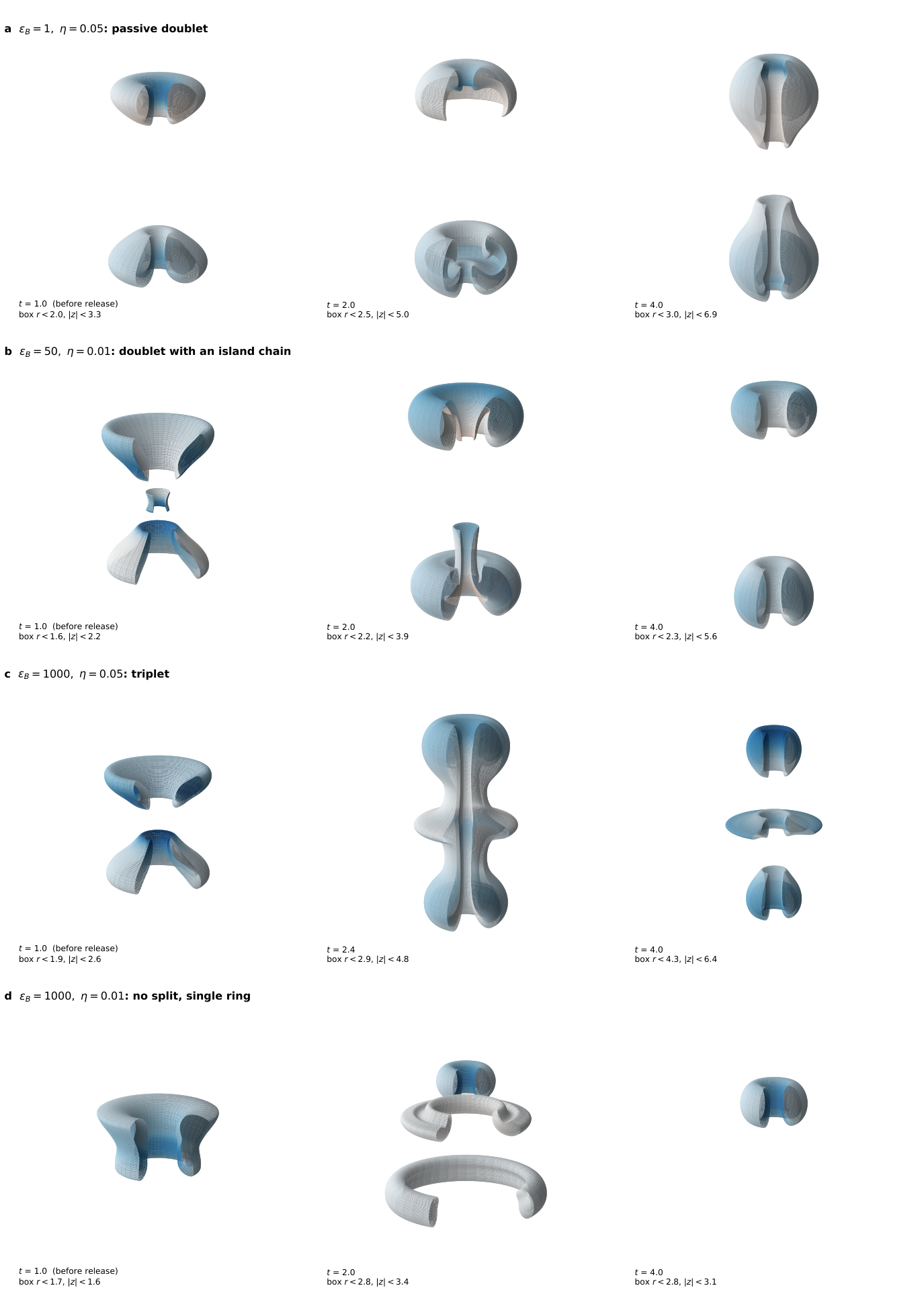}
\caption{\textbf{Four outcomes of the axisymmetric runs as revolved flux surfaces.} Flux surfaces revolved about the
axis and cut open, coloured by the toroidal field. Each run is shown at the release and at two later times, with the
window of each drawing given below it. \textbf{Panel a}, $\eB = 1$, $\eta = 0.05$, the passive doublet.
\textbf{Panel b}, $\eB = 50$, $\eta = 0.01$, the doublet with the island chain on the neck. \textbf{Panel c},
$\eB = 1000$, $\eta = 0.05$, the triplet. \textbf{Panel d}, $\eB = 1000$, $\eta = 0.01$, no split and a single
surviving ring. The large pale tori in Panel d at $t = 2$ are weak remnants.}
\label{fig:revolved}
\end{figure}

\paragraph{Parameters and results.} Supplementary Table~\ref{tab:common} lists the parameters common to all runs.
Supplementary Table~\ref{tab:results} lists the results of the eighteen runs of the scan. These are the data behind
Figs.~1, 3 and 4 of the main text. Supplementary Figs.~\ref{fig:snap005} and~\ref{fig:snap001} show every run at
$t = 1$, immediately before the release, and at $t = 1.5$, 2, 3 and 4. Supplementary Fig.~\ref{fig:revolved} shows
four of the outcomes as revolved flux surfaces. The energy ratio across the release is defined as in Fig.~4a of the
main text. It is the kinetic energy of the released flow at $t = 1.05$ divided by the magnetic energy at $t = 0.99$.
The Lundquist number is $S = \sqrt{\eB}\,B_{\max}/\eta$. Here $B_{\max}$ is the peak field at the release and the seed
radius is the length.

\begin{table}
\centering
\caption{\textbf{Parameters common to all runs.} The scan varies the magnetic coupling $\eB$ from 1 to 1000 at the
two magnetic diffusivities $\eta = 0.05$ and 0.01.}
\label{tab:common}
\small
\begin{tabular}{ll}
\toprule
box and lattice & $0 < r < 6$, $|z| < 8$, $384\times1024$ cells\\
imposed flow & reference branch of the reconstruction, $j = 0.001$, $\delta = 0.005$\\
 & column cutoff at $R_c = 16$, taper from 75\% to 90\% of the box\\
swirl & $A = 2$, $X_R = 8$, $X_c = 0.5$, counter-clockwise\\
regularization and release & $\tau_f = 0.05$, sudden release at $t = 1$, run to $t = 4$\\
seed & Taylor spheromak of radius 1, $B_\phi < 0$ above the midplane ($\sigma = -1$)\\
viscosity & $\nu = 0.1$\\
\bottomrule
\end{tabular}
\end{table}

\paragraph{Reading the outcomes.} Every 0.1 time units the local maxima of $\Psi$ in the meridional plane are found.
The depth of each maximum is its value minus the highest pass connecting it to a higher maximum. A ring is
\emph{separate} when it is deeper than 5\% of $\Psi_{\max}$. These are the ring records of Fig.~4b of the main text.
The string in Supplementary Table~\ref{tab:results} gives the number of rings every 0.3 time units. The outcome at
$t = 4$ in Fig.~4d of the main text is read from the number of separate rings at the end of the run. It is refined by
inspection of the snapshots and of the shallower islands on the neck. The transient episodes of three or four rings
at $\eta = 0.05$ and $20\le\eB\le300$ are rolls inside the daughters (Supplementary Fig.~\ref{fig:snap005}). The
persistent extra rings at $\eta = 0.01$ and $10\le\eB\le300$ are the chain of small islands on the neck between the
daughters (Supplementary Fig.~\ref{fig:snap001}). The three rings of $\eB = 1000$, $\eta = 0.05$ are the triplet. The
late third ring of $\eB = 500$, $\eta = 0.05$ is the central ring beginning to form. The asymmetric two-to-three-ring
evolution at $\eB = 500$, $\eta = 0.01$ is read directly from the ring history. So is the surviving single ring at
$\eB = 1000$, $\eta = 0.01$. That run does not split before the release. Its torus is compressed into a sheet and
develops lobes, but it is never pinched through. After the release the ring is displaced along the axis. It sheds
weak remnants and survives as a single ring (Supplementary Fig.~\ref{fig:revolved}d).

\begin{table}
\centering
\caption{\textbf{Results of the eighteen runs.} $t_s$ is the split time and ratio the island-depth ratio at the birth
(Fig.~1c,d of the main text). $t_2$ is the time at which the flux on the neck between the daughters falls below 25\% of
$\Psi_{\max}$. From then on the index sum read along the tracked ridge is 2 (Supplementary Note~\ref{sec:res}). $S$ is
the Lundquist number at the release. $H(1)/H(0)$ is the helicity kept through the collapse and $K/M$ the energy ratio
across the release. $H(4)/H(1)$ and $M(4)/M(1)$ are the helicity and the magnetic energy kept from the release to
$t = 4$. The string gives the number of rings every 0.3 time units from $t = 1$ to 4, counting the main ring plus the
separate secondary rings. The last column is the outcome of Fig.~4d of the main text. The $\eB = 1000$, $\eta = 0.01$
run does not split before the release.}
\label{tab:results}
\footnotesize\setlength{\tabcolsep}{3.4pt}
\begin{tabular}{rrcccrrrccll}
\toprule
$\eB$ & $\eta$ & $t_s$ & ratio & $t_2$ & $S$ & $\frac{H(1)}{H(0)}$ & $K/M$ & $\frac{H(4)}{H(1)}$ & $\frac{M(4)}{M(1)}$ & rings & outcome at $t = 4$\\
\midrule
1 & 0.05 & 0.305 & 0.76 & 0.71 & 5 & 0.093 & $4.4\times10^{4}$ & 0.056 & 0.088 & 22222222222 & passive doublet\\
3 & 0.05 & 0.308 & 0.58 & 0.72 & 8 & 0.091 & $1.6\times10^{4}$ & 0.058 & 0.093 & 22222222222 & passive doublet\\
10 & 0.05 & 0.320 & 0.36 & 0.74 & 13 & 0.089 & $5.1\times10^{3}$ & 0.066 & 0.105 & 22222222222 & passive doublet\\
20 & 0.05 & 0.339 & 0.26 & 0.76 & 17 & 0.091 & $2.5\times10^{3}$ & 0.074 & 0.119 & 22222422222 & doublet, rolls\\
50 & 0.05 & 0.416 & 0.19 & 0.82 & 25 & 0.100 & $8.6\times10^{2}$ & 0.089 & 0.147 & 22244322222 & doublet, rolls\\
100 & 0.05 & 0.545 & 0.20 & 0.87 & 37 & 0.108 & $3.6\times10^{2}$ & 0.101 & 0.165 & 22242222222 & doublet, rolls\\
200 & 0.05 & 0.657 & 0.10 & 0.92 & 140 & 0.114 & $1.6\times10^{2}$ & 0.106 & 0.155 & 22222222222 & doublet, rolls\\
300 & 0.05 & 0.708 & 0.09 & 0.95 & 221 & 0.118 & 94.7 & 0.102 & 0.128 & 24222222222 & doublet, rolls\\
500 & 0.05 & 0.778 & 0.09 & 0.98 & 164 & 0.123 & 50.6 & 0.086 & 0.089 & 24222222223 & doublet, central ring\\
1000 & 0.05 & 0.881 & 0.11 & 1.01 & 211 & 0.133 & 21.8 & 0.063 & 0.054 & 22222333333 & triplet\\
\midrule
1 & 0.01 & 0.348 & 0.78 & 0.80 & 140 & 0.444 & $4.5\times10^{3}$ & 0.123 & 0.419 & 22222222222 & passive doublet\\
10 & 0.01 & 0.378 & 0.46 & 0.87 & 1166 & 0.440 & $6.2\times10^{2}$ & 0.085 & 0.238 & 22244444433 & doublet, neck chain\\
50 & 0.01 & 0.703 & 0.18 & 1.01 & 1474 & 0.460 & 78 & 0.127 & 0.139 & 32222222222 & doublet, neck chain\\
100 & 0.01 & 0.921 & 0.29 & 1.03 & 2383 & 0.516 & 29.9 & 0.092 & 0.074 & 33322244444 & doublet, neck chain\\
200 & 0.01 & 0.959 & 0.58 & 1.05 & 3334 & 0.560 & 13.9 & 0.089 & 0.050 & 36655644433 & doublet, neck chain\\
300 & 0.01 & 0.955 & 0.56 & 1.08 & 3855 & 0.583 & 9.47 & 0.118 & 0.062 & 25443332322 & doublet, neck chain\\
500 & 0.01 & 0.998 & 0.38 & 1.06 & 4144 & 0.610 & 6.1 & 0.144 & 0.081 & 13333332222 & asymmetric, 2--3 rings\\
1000 & 0.01 & -- & -- & -- & 4277 & 0.640 & 3.63 & 0.282 & 0.172 & 12232222111 & single ring\\
\bottomrule
\end{tabular}
\end{table}

\section{The source of the asymmetry of the fold}
\label{sec:asym}
\paragraph{Two sources.} The model has two sources of up--down asymmetry. The offset $j = 0.001$ displaces the
stagnation plane of the imposed flow slightly from the midplane. The second source is the winding of the field by the
swirl. The flux function $\Psi$ of the seed and the angular velocity $\Omega_U$ of the swirl are even in $z$. The
winding rate $\mathbf B_p\cdot\nabla\Omega_U$, with $\mathbf B_p$ the poloidal field, is therefore odd in $z$. The
swirl thus adds an odd toroidal field to the even toroidal field of the seed. Near the axis it winds negative
toroidal field above the midplane and positive below it. In the outer part of the torus the signs are opposite. The
seed has $B_\phi = \sigma\lambda_T\Psi/r$, with $\sigma = -1$ in every run except the control with the twist
reversed. The wound field is therefore parallel to the twist in the upper half-space and opposed to it in the lower.
The fraction of $|B_\phi|$ opposite to the twist stays below 6\% up to the release in every run and below 9\%
afterwards. The one exception is $\eB = 1$, $\eta = 0.01$, where it reaches 24\% at $t = 4$. The toroidal field acts
on the meridional flow through the term $-\eB\,\partial_z(B_\phi^2/r^2)$ of the equation for the azimuthal vorticity
over $r$. The cross term in $B_\phi^2$ between the even and the odd parts is odd in $z$, so its $z$ derivative is
even. It drives an up--down antisymmetric meridional response. This response grows with $\eB$ and displaces the
pinch.

\begin{figure}[tbp]
\centering
\includegraphics[width=\textwidth]{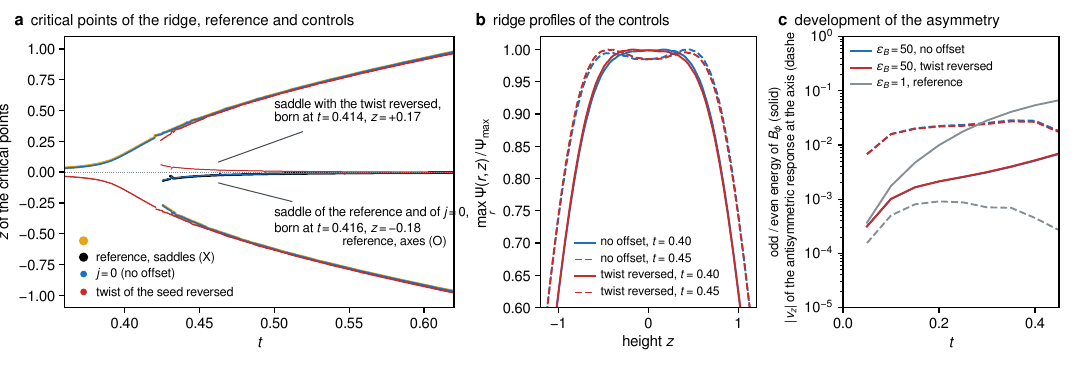}
\caption{\textbf{The source of the asymmetry of the fold at $\eB = 50$, $\eta = 0.05$.} \textbf{Panel a}, Heights of
the resolved critical points of the flux ridge against time. The reference run has its axes in gold and its saddles
in black. The control without the offset is blue and the control with the twist of the seed reversed is red.
\textbf{Panel b}, Ridge profiles of the two controls at $t = 0.40$ and 0.45. They are mirror images of each other.
\textbf{Panel c}, Energy of the odd part of $B_\phi$ relative to the even part (solid) and axial velocity of the
up--down antisymmetric response at the magnetic axis (dashed). The curves are for the two controls and for the
$\eB = 1$, $\eta = 0.05$ run.}
\label{fig:asym}
\end{figure}

\paragraph{Controls.} Two controls at $\eB = 50$, $\eta = 0.05$ separate the two sources (Supplementary
Table~\ref{tab:sym} and Supplementary Fig.~\ref{fig:asym}). Without the offset the split occurs 0.0005 earlier, with
the saddle at the same height. With the twist of the seed reversed the split is the mirror image. The pinch therefore
takes the side on which the wound field opposes the twist. The offset contributes at the 5\% level. In both controls
the odd part of $B_\phi$ carries 0.5\% of the energy of the even part at $t = 0.40$. The antisymmetric axial velocity
at the magnetic axis is 0.028. At $\eB = 1$ the odd part is larger, 5\%. The velocity it drives is
$4.5\times10^{-4}$, sixty times smaller, so the split stays close to a pitchfork. Without the offset and the swirl the
equations are symmetric about the midplane. The split at $\eB = 1$ is then an exact pitchfork, with the depth exponent
2.01. In all five runs of Supplementary Table~\ref{tab:sym} the median island-depth ratio lies within 0.02 of the cusp
prediction $R(\beta)$ of Supplementary Note~\ref{sec:cusp}.

\begin{table}
\centering
\caption{\textbf{Symmetry controls.} The columns give the offset $j$ of the stagnation plane and the sign $\sigma$ of the
twist of the seed. They then give the split time $t_s$ and the height of the saddle at its first detection. The ratio is the
island-depth ratio when the new island first exceeds $10^{-3}\Psi_{\max}$. The medians of $\beta$ and of the ratio
are taken over the first fifteen records deeper than $5\times10^{-4}\Psi_{\max}$. $R(\beta)$ is the cusp prediction
for that $\beta$ (Supplementary Note~\ref{sec:cusp}). Upper block, $\eB = 50$, $\eta = 0.05$. Lower block, $\eB = 1$,
$\eta = 0.05$, with the exactly symmetric run.}
\label{tab:sym}
\small
\begin{tabular}{lcccccccc}
\toprule
run & $j$ & $\sigma$ & $t_s$ & $z$ of the saddle & ratio & $\beta$ & ratio (median) & $R(\beta)$\\
\midrule
reference & 0.001 & $-1$ & 0.4160 & $-0.183$ & 0.19 & 0.199 & 0.192 & 0.172\\
no offset & 0 & $-1$ & 0.4155 & $-0.183$ & 0.21 & 0.189 & 0.208 & 0.192\\
twist reversed & 0.001 & $+1$ & 0.4142 & $+0.167$ & 0.21 & 0.181 & 0.227 & 0.208\\
\midrule
reference, $\eB = 1$ & 0.001 & $-1$ & 0.3054 & $-0.073$ & 0.76 & 0.031 & 0.773 & 0.780\\
no offset, no swirl, $\eB = 1$ & 0 & $-1$ & 0.3037 & $0.000$ & 1.00 & 0.000 & 1.000 & 1.000\\
\bottomrule
\end{tabular}
\end{table}

\section{Lattice dependence of the local observables}
\label{sec:res}
\paragraph{A fold of the rounded torus.} The fold of the $\eB = 50$, $\eta = 0.05$ run is shown in Fig.~2b of the main
text. This run was repeated on a $512\times1365$ lattice of the production box to $t = 0.5$. The energies, the helicity and the
peak field agree with the production lattice to 0.05\%. Supplementary Table~\ref{tab:res} compares the birth of the new
island on the two lattices. The first detection of the island, which defines $t_s$, is set by the noise floor of the
ridge. It comes at $t = 0.4160$ and a depth of $2\times10^{-5}\Psi_{\max}$ on the production lattice. On the finer
lattice it comes at 0.4186 and $4\times10^{-5}\Psi_{\max}$. The split time is therefore uncertain by about 0.003.
From a depth of $10^{-4}\Psi_{\max}$ on, the island reaches every depth up to $10^{-2}\Psi_{\max}$ within 0.0004 of
the same time. The saddle is then within 0.005 of the same height. The positions of the critical points, the
island-depth ratio, the fitted birth time and the growth exponents agree as well. On both lattices the median ratio
lies within 0.02 of $R(\beta)$.

\paragraph{A late split of the flat sheet.} The $\eB = 100$, $\eta = 0.01$ run was repeated on the same lattice to
$t = 4$. Its integral quantities and its energy ratio across the release agree with the production lattice to a few
per cent. Its ring history is the same up to one more island in the neck chain (Supplementary Table~\ref{tab:res}).
The split itself comes later and more central. The late splits at $\eta = 0.01$ happen on a flat top. The top breaks
into several shallow maxima before the first resolved island appears (Supplementary Fig.~\ref{fig:fit}d--f of
Supplementary Note~\ref{sec:cusp}). Small perturbations decide which of them grows. These splits are therefore
qualitatively robust but not converged in their timing and symmetry.

\begin{table}[tbp]
\centering
\caption{\textbf{Lattice dependence of two splits.} Upper block, the fold of the rounded torus at $\eB = 50$,
$\eta = 0.05$, computed to $t = 0.5$. Depths are in units of $\Psi_{\max}$. The first detection is given with its
depth and the height of the saddle. The next rows give the times at which the new island reaches a depth, with the
height of the saddle. The cusp statistics are defined as in Supplementary Table~\ref{tab:sym}. Lower block, the late
split of the flat sheet at $\eB = 100$, $\eta = 0.01$, computed to $t = 4$.}
\label{tab:res}
\small\setlength{\tabcolsep}{4.5pt}
\begin{tabular}{lcc}
\toprule
 & $384\times1024$ & $512\times1365$\\
\midrule
\multicolumn{3}{l}{\emph{Fold, $\eB = 50$, $\eta = 0.05$}}\\
first detection of the new island, $t$ (depth, $z$ of the saddle) & 0.4160 ($2\times10^{-5}$, $-0.183$) & 0.4186 ($4\times10^{-5}$, $-0.117$)\\
depth $10^{-4}$ reached, $t$ ($z$ of the saddle) & 0.4194 ($-0.108$) & 0.4196 ($-0.103$)\\
depth $10^{-3}$ reached, $t$ ($z$ of the saddle) & 0.4257 ($-0.066$) & 0.4256 ($-0.065$)\\
depth $10^{-2}$ reached, $t$ ($z$ of the saddle) & 0.4499 ($-0.028$) & 0.4497 ($-0.029$)\\
$z$ of the three critical points at depth $3\times10^{-3}$ & $-0.325$, $-0.047$, 0.357 & $-0.324$, $-0.048$, 0.360\\
island-depth ratio at depth $10^{-3}$ & 0.192 & 0.192\\
median $\beta$, median ratio and $R(\beta)$ & 0.199, 0.192, 0.172 & 0.194, 0.168, 0.181\\
fitted birth time $t_b$ & 0.4175 & 0.4178\\
exponents of the depth and of the separation & 1.58, 0.49 & 1.60, 0.47\\
\midrule
\multicolumn{3}{l}{\emph{Late split, $\eB = 100$, $\eta = 0.01$}}\\
split time $t_s$ & 0.921 & 0.954\\
island-depth ratio at the birth & 0.29 & 0.50\\
Lundquist number at the release & 2380 & 2530\\
helicity kept to the release & 0.516 & 0.516\\
energy ratio across the release & 29.9 & 29.2\\
helicity kept, $t = 1\to4$ & 0.092 & 0.089\\
magnetic energy kept, $t = 1\to4$ & 0.074 & 0.071\\
separate rings, $t = 1.0$--1.8 / 1.9--2.5 / 2.6--4.0 & 2 / 1 / 2--3 & 2 / 1 / 2--4\\
\bottomrule
\end{tabular}
\end{table}

\paragraph{Bookkeeping of the critical points.} The O points of the flux ridge are its local maxima, the magnetic
axes. The X points are its local minima, the saddles. Both are found above 25\% of $\Psi_{\max}$. The index sum is
the number of O minus the number of X. For isolated critical points, none of which crosses the 25\% level, the
Poincar\'e--Hopf theorem fixes it. The sum equals 1 as long as the tracked part of the ridge contains all the saddles. Every
change in the number of critical points is then the creation or the annihilation of an O--X pair. The counts are
evaluated on all stored critical points, before the unresolved pairs are removed. The sum changes only when a critical
point crosses the 25\% level. This happens when the flux on the neck between separated rings falls below that level,
at the time $t_2$ of Supplementary Table~\ref{tab:results}. The saddle between the rings then leaves the tracked
region, and the sum read on the remainder rises by one. At $\eta = 0.01$ it also happens when a weak axis on a ring
already disconnected fades through that level. The eighteen runs contain 86 changes of the sum. Every one of them is
at a record in which a critical point lies within 0.007 of the 25\% level.

\section{The imposed flow and its regularization}
\label{sec:flow}
\paragraph{What the imposed flow retains from the construction.} In the notation of the Methods, the imposed flow
retains four elements of the reconstruction\cite{leiren2026} of the reported solution\cite{openai2026}. The first is the
similarity coordinates. The second is the axial velocity $(4Z + j)/\lambda^{1+\delta}$ of the axis, extended
unchanged across the column. The third is the incompressible radial flow following from it. The fourth is a swirl
with the structure of the reference branch. The imposed flow omits the construction's core corrections, connection region, shear modification, moment corrections, pressure,
admissible stress, oscillatory pulses and outer continuation. The singularity itself is regularized by $\tau_f$. The
swirl amplitude $A = 2$ is a model choice. Its order follows from equating the pressure deficit of the swirl on the
axis, about $2.5A^2$, to the dynamic pressure of the jets. The uniform axial jet and the $R^{1/10}$ swirl have the
shear parameter $\kappa = \tfrac45R/(R + X_c) + \tfrac65R^2/(X_R^2 + R^2) < 2$. They therefore do not satisfy the
admissible-stress condition of the full construction. The prescribed-flow equations do not require it.

\paragraph{The cutoff.} The meridional flow is switched off smoothly between the similarity radii $R_c = 16$ and
$2R_c$, measured with the midplane scale as $r^2/(2\tau_e)$. This leaves the flow unchanged inside a column of radius
$\sqrt{2R_c\tau_e}$ around the axis, 1.26 at $t = 1$. It switches the flow off by $\sqrt{4R_c\tau_e}$, 1.79 at
$t = 1$, with the return flow in between. The local similarity radius $r^2/(2\lambda(z)^2)$ would instead let the
flow fill a cone $r < \sqrt{2R_c}\,|z|$. In this cone the radial flow is outward, and the energy grows like $1/(1-t)$
and is set by the box. Control simulations with this cone cutoff were made at $\eB$ up to 500 and $\nu$ up to 1. The
split is unchanged. The response, however, is amplified a hundredfold just before $t = 1$. The released flow then
hurls the field into the walls and shreds it. The cone hands over a kinetic energy of $1.4\times10^5$ at the release,
against $2.4\times10^3$ for the column. The peak meridional speeds are 202 and 54.

\paragraph{The regularization.} With $\tau_e$ in place of $1 - t$, the poloidal energy of the imposed flow rises from
$5.0\times10^3$ at $t = 0$ to $8.0\times10^3$ near $t = 0.45$. It then falls as $\tau_e^{1/2}$, in line with Leray's
law\cite{leray1934}, to $2.1\times10^3$ at $t = 1$. The swirl energy grows from $1.3\times10^2$ to $2.3\times10^2$ and
saturates, because the outer swirl is a potential vortex. Supplementary Table~\ref{tab:tauf} gives the frozen flow at
$t = 1$ for $\tau_f$ from 0.4 to 0.0125. The poloidal energy at $t = 1$ scales as $\tau_f^{1/2}$, the peak speed as
$\tau_f^{-1/2}$ and the strain as $2/\tau_f$. These are Leray's exponents. The $L_2$ norm of the jump at the release
scales as $\tau_f^{-3/4}$ and its pointwise maximum as $\tau_f^{-3/2}$. In the limit $\tau_f\to0$ the poloidal energy
handed over vanishes. What remains is the swirl's vortex tail and the response the field has driven. At
$\tau_f = 0.05$ the imposed flow at the release carries 43\% of its initial poloidal energy, $2.4\times10^3$ in all.
The total kinetic energy of the released state, with the response, lies between $2.3\times10^3$ and $2.7\times10^3$
across the scan. The energy handed over, divided by the magnetic energy at the release, gives the energy ratios of
Fig.~4 of the main text. The only difference is its viscous decay by $t = 1.05$. These ratios order the regimes. Their dependence on
$\tau_f$ is therefore a dependence on how far the collapse is allowed to proceed.

\begin{table}[tbp]
\centering
\caption{\textbf{The frozen imposed flow at $t = 1$ against the regularization time.} Production lattice, column cutoff
$R_c = 16$. $K_{\rm pol}$ and $K_{\rm swirl}$ are the poloidal and swirl energies and $\gamma$ the strain at the
origin. $\|[\partial_t\mathbf u]\|_2$ and $\max|[\partial_t\mathbf u]|$ are the $L_2$ norm and the pointwise maximum
of the acceleration jump at the release. $T_{\rm jump}$ is its time scale. $K/D_\nu$ is the viscous decay time of
the imposed flow. The last column is the number of cells across the solid-body core of the swirl.}
\label{tab:tauf}
\small\setlength{\tabcolsep}{4.5pt}
\begin{tabular}{cccccccccc}
\toprule
$\tau_f$ & $K_{\rm pol}$ & $K_{\rm swirl}$ & $\max|\mathbf U|$ & $\gamma$ & $\|[\partial_t\mathbf
u]\|_2$ & $\max|[\partial_t\mathbf u]|$ & $T_{\rm jump}$ & $K/D_\nu$ & core cells\\
\midrule
0.4 & 5858 & 189 & 19.3 & 5.0 & 746 & 160 & 0.147 & 0.535 & 40\\
0.2 & 4181 & 216 & 27.2 & 10.0 & 1246 & 454 & 0.0753 & 0.284 & 29\\
0.1 & 2991 & 228 & 38.4 & 19.9 & 2093 & 1279 & 0.0383 & 0.147 & 20\\
0.05 & 2133 & 231 & 54.0 & 39.7 & 3507 & 3588 & 0.0196 & 0.077 & 14\\
0.025 & 1514 & 228 & 75.8 & 78.7 & 5851 & 9989 & 0.0101 & 0.041 & 10\\
0.0125 & 1068 & 222 & 105.3 & 154.9 & 9689 & 27292 & 0.00524 & 0.023 & 7\\
\bottomrule
\end{tabular}
\end{table}

\section{The release and the budgets}
\label{sec:release}
\paragraph{The jump at the release.} The release removes the force $\mathbf f$ of equation (1) of the main text,
which holds the imposed flow. The velocity and the field are continuous across $t = 1$, and so is
$\partial_t\mathbf B$. The acceleration jumps by $[\partial_t\mathbf u] = -\mathcal P\mathbf f$, with $\mathcal P$
the projection onto divergence-free fields. The field therefore feels the jump only at second order in $t - 1$. The
slope of the total energy jumps by $-P_{\rm ext}^-$, the power of the force just before the release. Because
$\dot\tau_e = -(1 - t)/\tau_e$ vanishes at $t = 1$, the imposed flow is frozen there. The jump is then
\begin{equation}
[\partial_t\mathbf u] = \mathcal P(\mathbf U\times\boldsymbol\omega_U) + \nu\nabla^2\mathbf U,
\label{eq:helm}
\end{equation}
the Helmholtz rate of the frozen vorticity $\boldsymbol\omega_U$. It injects no energy. An irrotational imposed flow
would produce no jump. At $\tau_f = 0.05$ the jump has $L_2$ norm 3507 and maximum 3588. Its inertial part is 3311
and its viscous part 851. The inertial part is close to $\gamma\|\mathbf U\|_2 = 2.7\times10^3$. This is the
acceleration produced by the strain $\gamma = 39.7$ in a flow of norm $\|\mathbf U\|_2 = 69$. The time scale of the
jump, $\|\mathbf U\|_2/\|[\partial_t\mathbf u]\|_2 = 0.020$, is $0.8/\gamma$. It lies between $0.74/\gamma$ and
$0.81/\gamma$ for every $\tau_f$ of Supplementary Table~\ref{tab:tauf}. Relative to the strain the jump therefore does
not depend on the regularization. The $L_2$ norm of $\partial_t\mathbf U$ is $3.0\times10^3$ at $t = 0.90$,
$3.4\times10^3$ at 0.93--0.95 and $2.2\times10^3$ at 0.98. The release thus restores an acceleration of the size the
collapse had at $t\approx0.93$--0.97. At $t = 1$ the net power of the force on the frozen imposed flow equals its
viscous dissipation, $3.1\times10^4$. The inertial part does no net work on $\mathbf U$. Its power on the
total velocity, $2.7$--$3.5\times10^4$, lies within 13\% of this value in every run. The regularization thus smooths
the collapse and not the release.

\paragraph{Energy and helicity budgets.} The total velocity is $\mathbf u = \mathbf U + \mathbf v$. The energies
and the helicity, their budgets and the rates entering them are collected below.
\begin{center}
\small\renewcommand{\arraystretch}{1.08}
\begin{tabular}{@{}ll@{\hspace{2.5em}}ll@{}}
\toprule
energies and helicity & budgets & rates & \\
\cmidrule(r){1-2}\cmidrule(l){3-4}
$K = \tfrac12\!\int|\mathbf u|^2\,dV$ & $\dot K = W_L - D_\nu + P_{\rm ext}$ &
$W_L = \eB\!\int\mathbf u\cdot(\mathbf J\times\mathbf B)\,dV$ & power of the Lorentz force\\
$M = \tfrac{\eB}2\!\int|\mathbf B|^2\,dV$ & $\dot M = -W_L - D_\eta$ &
$D_\nu = \nu\!\int|\nabla\mathbf u|^2\,dV$ & viscous dissipation\\
$E = K + M$ & $\dot E = P_{\rm ext} - D_\nu - D_\eta$ &
$D_\eta = \eB\eta\!\int|\mathbf J|^2\,dV$ & resistive dissipation\\
$H = \int\mathbf A\cdot\mathbf B\,dV = 4\pi\!\int\Psi B_\phi\,dr\,dz$ & $\dot H = -2\eta\!\int\mathbf J\cdot\mathbf B\,dV$ &
$P_{\rm ext} = \int\mathbf u\cdot\mathbf f\,dV$ & power of the external force\\
\bottomrule
\end{tabular}
\end{center}
The force acts only up to the release. Afterwards $E(t) + \int_1^t(D_\nu + D_\eta)\,dt' = E(1)$. The boundary fluxes
vanish for the tapered background and the compact field. Because $\Psi$ vanishes on the axis and on the walls, $H$ is
gauge invariant. Its budget holds before and after the release.

\begin{figure}[tbp]
\centering
\includegraphics[width=\textwidth]{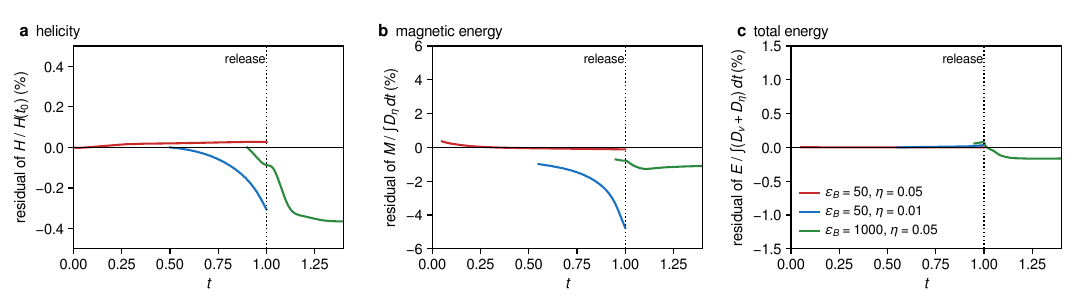}
\caption{\textbf{Closure of the budgets in time.} The residual of a budget is the stored quantity minus the time
integral of its rates. It is shown for the three re-integrated runs. Red, $\eB = 50$, $\eta = 0.05$ from $t = 0$. Blue,
$\eB = 50$, $\eta = 0.01$ from $t = 0.5$. Green, $\eB = 1000$, $\eta = 0.05$ from $t = 0.9$ to 1.4, across the
release. The dotted line marks the release. \textbf{Panel a}, Residual of the helicity relative to the
helicity at the start of the interval. \textbf{Panel b}, Residual of the magnetic energy relative to the resistive
dissipation accumulated since the start. It is shown from 0.05 time units after the start. \textbf{Panel c}, Residual
of the total energy relative to the energy dissipated since the start, with the quadrature imbalance of the analytic
imposed flow removed.}
\label{fig:budget}
\end{figure}

\paragraph{Closure of the budgets in time.} Three runs were re-integrated with the production code. The rates of the
table were evaluated every fourth step from the same discrete fields and operators as the energies. The first run is
the fold at $\eB = 50$, $\eta = 0.05$, from $t = 0$ to the release. The second is the same coupling at $\eta = 0.01$,
from $t = 0.5$ to the release. The third is the triplet at $\eB = 1000$, $\eta = 0.05$, from $t = 0.9$ to 1.4,
across the release. The residual
of a budget is the stored quantity minus the time integral of its rates. Supplementary Fig.~\ref{fig:budget} shows the
residuals. For the helicity the residual stays below 0.03\%, 0.3\% and 0.4\% of the initial helicity in the three
runs. It stays below 0.8\% of the helicity dissipated. For the magnetic energy the residual stays below 0.1\% and
1.3\% of the resistive dissipation at $\eta = 0.05$. It reaches 4.7\% at $\eta = 0.01$, where the resistive layer
spans one to one and a half cells. Before the release the residual of the kinetic budget equals the discrete
imbalance of the analytic imposed flow alone, $\int\mathbf U\cdot\mathbf f\,dV - \dot K_U - D_\nu(\mathbf U)$. The
two agree to within 0.1\% of the dissipated energy. This imbalance is a quadrature error of the second-order differences. In it the
power of the force on the imposed flow and the viscous dissipation of the imposed flow, each about $2\times10^4$,
nearly cancel. After the release no imposed flow enters. There $E(t) - E(1) + \int_1^t(D_\nu + D_\eta)\,dt'$ is $-10$
at $t = 1.4$, 0.5\% of the energy dissipated. The accumulated error of the discretization includes the numerical
dissipation of the upwind advection. It is therefore half a per cent of the physical dissipation in the most violent
interval of the scan.

\paragraph{Smooth switch-off of the force.} The smooth release sets $\mathbf U = s(t)\mathbf U_0(t)$. Here
$s = C((t - 1)/\Delta T)$ is the smooth step of the Methods, with $\Delta T = 0.05$. It falls from 1 at $t = 1$ to 0
at $t = 1 + \Delta T$. The transfer $-\dot s\,\mathbf U_0$ is added to the response at every Runge--Kutta stage.
$\mathbf U_0$ keeps the time dependence of the regularized profile, in which $\tau_e$ grows again after $t = 1$. The
profile therefore re-expands slowly during the ramp. The force is then
$s\,\partial_t\mathbf U_0 + s^2(\mathbf U_0\cdot\nabla)\mathbf U_0 - \nu s\nabla^2\mathbf U_0$ up to a gradient. Its
power on $\mathbf u$ is the $P_{\rm ext}$ of the budgets. Its work $W$ is the integral of that power over the ramp.
We tested it at $\eB = 1000$, $\eta = 0.05$, the run producing the triplet. Restarted from its snapshot at $t = 0.9$
with the sudden release, the run reproduces the original to every printed digit. The twenty-three pair events up to
$t = 2.5$ are reproduced to four decimals. The force does work $W = 491$ during the ramp, 21\% of the kinetic energy
at the release. The surplus is dissipated on the viscous decay time of the released flow, 0.075. The compression of
the field after the release peaks at $M = 89$ at $t = 1.124$ instead of 95 at 1.083. The kinetic energy differs by
12--18\% during the first half time unit and by less than 6\% afterwards. The magnetic energy differs by up to 45\%
at $t = 1.2$ and by 13--20\% from $t = 3$ to 4. The helicity differs by 11--13\% from $t = 1.5$ to 2.5 and within 2\%
at $t = 4$. The pair events after the release keep their kinds and order. The lobe islands of the daughters are born
within 0.011 of the same times. They persist to $t = 1.61$--1.65 instead of 1.39--1.41. The saddle between the
daughters appears at $t = 1.95$ instead of 1.70. The outer rings are shed in folds at $t = 2.08$ and 2.09 instead of
1.81 and 2.03. One further transient pair appears on a lobe. The inner cores merge at $t = 2.65$--2.70 instead of
2.28. The triplet persists to $t = 4$ in both runs. These numbers bound the influence of the protocol on the timing
of the events after the release, in the case most sensitive to it.

\section{Scope and robustness of the cusp relation}
\label{sec:cusp}
\paragraph{The asymmetry from the positions.} The three critical points $z_1 < z_2 < z_3$ of the ridge are shifted
so as to make their mean zero. They then give $a_{\rm pos} = -(z_1z_2 + z_1z_3 + z_2z_3)$ and
$b_{\rm pos} = z_1z_2z_3$, the coefficients of $x^3 - ax - b$ in position units. The combination
$\beta = |b_{\rm pos}|/a_{\rm pos}^{3/2}$ does not change under the rescaling of the normal form. It therefore follows
from the positions alone, independently of the island depths measured from the flux values of the same points. The
prediction $R(\beta)$ is evaluated from the roots $x_1 < x_2 < x_3$ of $x^3 - x - \beta = 0$. With
$f(x) = -x^4/4 + x^2/2 + \beta x$, it is the smaller of $f(x_1) - f(x_2)$ and $f(x_3) - f(x_2)$ divided by the larger.

\paragraph{Thresholds and window.} Supplementary Table~\ref{tab:cusp} gives, for every split, the deviation of the
measured ratio from $R(\beta)$. The depth thresholds are $2\times10^{-4}$, $5\times10^{-4}$ and $10^{-3}\Psi_{\max}$
and the windows 10, 15 and 25 records. For the eleven single-island births the deviation stays below 0.09 in every
combination with the two larger thresholds. The one exception is $\eB = 500$, $\eta = 0.05$, where it reaches 0.13
with the longer windows. There the new island grows quickly away from the birth. With the smallest threshold the two
single-island births at $\eta = 0.01$ scatter to deviations of 0.2. Their ridges carry ripples near that level. Six
births are on rippled ridges. The four at $\eta = 0.01$ and $\eB = 50$--300 lie 0.2--0.4 above the curve for every
combination with the two larger thresholds. The one at $\eB = 1000$, $\eta = 0.05$ lies 0.1 below it. The one at
$\eB = 500$, $\eta = 0.01$ falls on it, with the widest spread of the scan.

\begin{table}[tbp]
\centering
\caption{\textbf{The cusp test and its sensitivity.} For every split, the median asymmetry $\beta$, the median
island-depth ratio and the prediction $R(\beta)$ are given for the standard choice, a threshold of
$5\times10^{-4}\Psi_{\max}$ and fifteen records. The following columns give the deviation of the median ratio from
$R(\beta)$. They cover the nine combinations of the depth threshold $d_{\min}$ and the number $k$ of records in the
window.
Runs marked $\dagger$ are births on a ridge already carrying several shallow islands, which a single quartic does not
describe.}
\label{tab:cusp}
\footnotesize
\setlength{\tabcolsep}{3.4pt}
\renewcommand{\arraystretch}{1.0}
\begin{tabular}{rrccc|ccc|ccc|ccc}
\toprule
 & & & & & \multicolumn{3}{c|}{$d_{\min} = 2\times10^{-4}$} & \multicolumn{3}{c|}{$d_{\min} = 5\times10^{-4}$} & \multicolumn{3}{c}{$d_{\min} = 10^{-3}$}\\
$\eB$ & $\eta$ & $\beta$ & ratio & $R(\beta)$ & $k = 10$ & 15 & 25 & $k = 10$ & 15 & 25 & $k = 10$ & 15 & 25\\
\midrule
1 & 0.05 & 0.031 & 0.773 & 0.780 & $-0.02$ & $-0.01$ & $-0.00$ & $-0.01$ & $-0.01$ & $-0.00$ & $-0.00$ & $-0.00$ & $-0.00$\\
3 & 0.05 & 0.071 & 0.594 & 0.561 & $-0.01$ & $+0.04$ & $+0.03$ & $+0.05$ & $+0.03$ & $-0.01$ & $-0.01$ & $-0.01$ & $-0.00$\\
10 & 0.05 & 0.116 & 0.382 & 0.384 & $-0.00$ & $-0.00$ & $-0.00$ & $-0.00$ & $-0.00$ & $-0.00$ & $-0.00$ & $-0.00$ & $-0.00$\\
20 & 0.05 & 0.147 & 0.275 & 0.289 & $-0.00$ & $-0.02$ & $-0.01$ & $-0.01$ & $-0.01$ & $+0.00$ & $-0.01$ & $+0.01$ & $+0.01$\\
50 & 0.05 & 0.199 & 0.192 & 0.172 & $+0.02$ & $+0.02$ & $+0.02$ & $+0.02$ & $+0.02$ & $+0.01$ & $+0.01$ & $+0.01$ & $-0.04$\\
100 & 0.05 & 0.162 & 0.251 & 0.251 & $-0.06$ & $-0.02$ & $+0.04$ & $-0.02$ & $+0.00$ & $+0.05$ & $+0.02$ & $+0.05$ & $+0.08$\\
200 & 0.05 & 0.242 & 0.136 & 0.103 & $+0.01$ & $+0.02$ & $-0.03$ & $+0.03$ & $+0.03$ & $-0.01$ & $-0.03$ & $-0.02$ & $+0.01$\\
300 & 0.05 & 0.256 & 0.112 & 0.085 & $+0.04$ & $+0.04$ & $+0.04$ & $+0.04$ & $+0.03$ & $+0.03$ & $+0.03$ & $+0.03$ & $+0.06$\\
500 & 0.05 & 0.192 & 0.265 & 0.186 & $+0.06$ & $+0.06$ & $+0.09$ & $+0.07$ & $+0.08$ & $+0.12$ & $+0.07$ & $+0.11$ & $+0.13$\\
1000$^\dagger$ & 0.05 & 0.169 & 0.121 & 0.235 & $+0.03$ & $-0.11$ & $-0.11$ & $-0.03$ & $-0.11$ & $-0.12$ & $-0.11$ & $-0.11$ & $-0.12$\\
1 & 0.01 & 0.029 & 0.824 & 0.791 & $+0.18$ & $+0.23$ & $+0.02$ & $+0.02$ & $+0.03$ & $+0.03$ & $+0.03$ & $+0.04$ & $+0.03$\\
10 & 0.01 & 0.100 & 0.490 & 0.441 & $+0.17$ & $+0.16$ & $+0.16$ & $+0.07$ & $+0.05$ & $+0.07$ & $+0.08$ & $+0.06$ & $+0.07$\\
50$^\dagger$ & 0.01 & 0.368 & 0.224 & 0.003 & $+0.12$ & $+0.13$ & $+0.22$ & $+0.22$ & $+0.22$ & $+0.22$ & $+0.28$ & $+0.28$ & $+0.37$\\
100$^\dagger$ & 0.01 & 0.116 & 0.738 & 0.383 & $+0.37$ & $+0.35$ & $+0.35$ & $+0.37$ & $+0.35$ & $+0.35$ & $+0.37$ & $+0.35$ & $+0.35$\\
200$^\dagger$ & 0.01 & 0.109 & 0.746 & 0.409 & $+0.33$ & $+0.34$ & $+0.34$ & $+0.33$ & $+0.34$ & $+0.34$ & $+0.33$ & $+0.34$ & $+0.34$\\
300$^\dagger$ & 0.01 & 0.344 & 0.289 & 0.012 & $+0.01$ & $+0.23$ & $+0.25$ & $+0.26$ & $+0.28$ & $+0.30$ & $+0.38$ & $+0.37$ & $+0.36$\\
500$^\dagger$ & 0.01 & 0.113 & 0.378 & 0.395 & $+0.04$ & $-0.02$ & $-0.03$ & $+0.04$ & $-0.02$ & $-0.03$ & $+0.04$ & $-0.02$ & $-0.03$\\
\bottomrule
\end{tabular}
\end{table}

\paragraph{Birth times and growth exponents.} The birth times of the main text are linear fits of the squared
axis--saddle separation. This presumes the square-root law. Supplementary Table~\ref{tab:births} compares them, for
the nine single-island births at $\eta = 0.05$, with free fits. In the free fits the birth time, the exponent and the
prefactor are all free. They are made separately for the depth and for the separation, over the first 0.05 time units
after the birth. The free birth times differ from the two-step ones by at most 0.0022. The free exponents of the
separation lie between 0.42 and 0.58. The exponent of the depth is more sensitive to the birth time. It runs from
1.91 to 1.34 in the two-step fits and from 1.64 to 1.85 in the free fits. At $\eta = 0.01$ the top of the ridge
carries ripples at the $10^{-4}$ level at the birth. The growth fits are then not stable against the threshold and
the window, so no exponents are quoted for those runs.

\begin{table}[tbp]
\centering
\caption{\textbf{Birth times and growth exponents of the single-island splits at $\eta = 0.05$.} The split time $t_s$,
the birth time $t_b$ and the exponents of the island depth and of the axis--saddle separation. The two-step fits are
those of the main text. The free fits determine $t_b$, the exponent and the prefactor together, separately for the
depth and for the separation, over $0 < t - t_b < 0.05$. $n$ is the number of resolved records in the window. The
normal form gives 2 and 1/2 for the pitchfork and 3/2 and 1/2 for the fold.}
\label{tab:births}
\small\setlength{\tabcolsep}{4.2pt}
\begin{tabular}{rccccccccc}
\toprule
 & & \multicolumn{3}{c}{two-step fits} & \multicolumn{2}{c}{free fit, depth} & \multicolumn{2}{c}{free fit, separation} & \\
\cmidrule(lr){3-5}\cmidrule(lr){6-7}\cmidrule(lr){8-9}
$\eB$ & $t_s$ & $t_b$ & depth & separation & $t_b$ & exponent & $t_b$ & exponent & $n$\\
\midrule
1 & 0.3054 & 0.3043 & 1.91 & 0.51 & 0.3046 & 1.85 & 0.3054 & 0.43 & 101\\
3 & 0.3083 & 0.3075 & 1.75 & 0.49 & 0.3072 & 1.81 & 0.3076 & 0.48 & 102\\
10 & 0.3196 & 0.3184 & 1.74 & 0.49 & 0.3185 & 1.72 & 0.3188 & 0.47 & 102\\
20 & 0.3395 & 0.3385 & 1.67 & 0.49 & 0.3385 & 1.67 & 0.3390 & 0.46 & 104\\
50 & 0.4160 & 0.4175 & 1.58 & 0.49 & 0.4168 & 1.72 & 0.4177 & 0.48 & 110\\
100 & 0.5448 & 0.5481 & 1.70 & 0.52 & 0.5483 & 1.66 & 0.5496 & 0.42 & 113\\
200 & 0.6572 & 0.6614 & 1.48 & 0.47 & 0.6606 & 1.64 & 0.6620 & 0.42 & 124\\
300 & 0.7083 & 0.7122 & 1.43 & 0.46 & 0.7107 & 1.70 & 0.7121 & 0.47 & 124\\
500 & 0.7783 & 0.7835 & 1.34 & 0.46 & 0.7813 & 1.82 & 0.7820 & 0.58 & 139\\
\bottomrule
\end{tabular}
\end{table}

\begin{figure}
\centering
\includegraphics[width=0.8\textwidth]{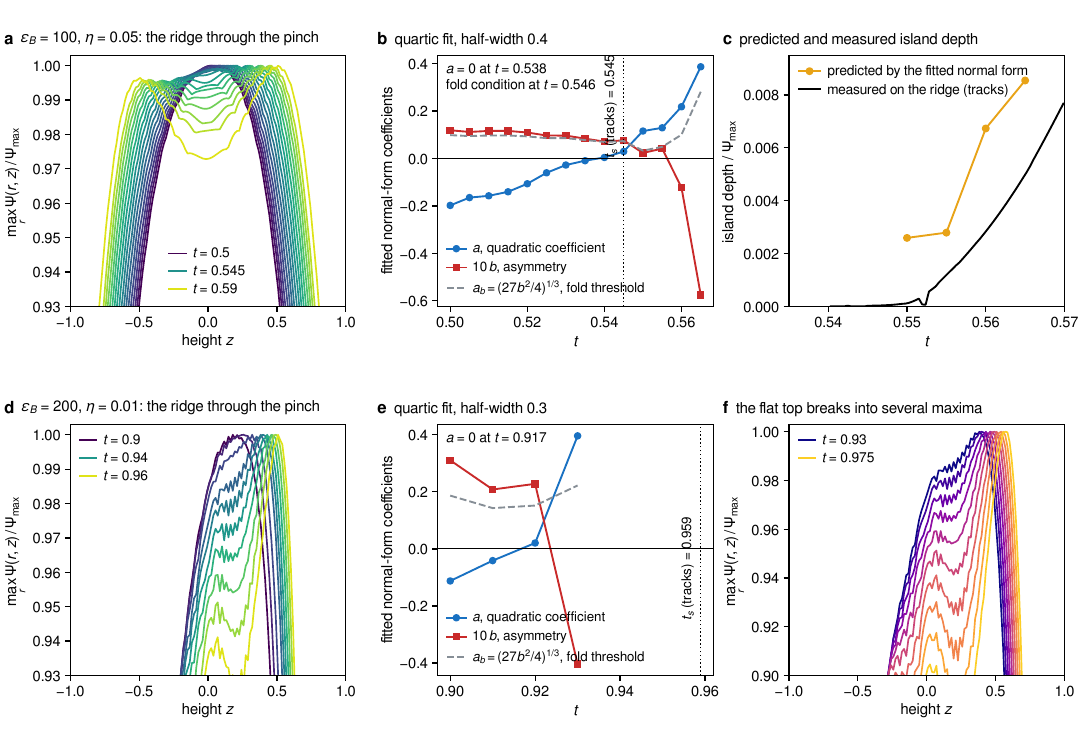}
\caption{\textbf{Direct fit of the flux ridge to the cusp normal form through the pinch.} Snapshots every 0.005. The
ridge is smoothed over two cells and fitted about its top within a half-width of 0.4 (Panels a--c) and 0.3 (Panels
d--f). Panels a--c, The fold at $\eB = 100$, $\eta = 0.05$. \textbf{Panel a}, Ridge profiles from $t = 0.5$ to 0.59.
\textbf{Panel b}, Fitted quadratic coefficient $a$ and asymmetry $b$ (shown as $10b$), with the fold threshold
$a_b = (27b^2/4)^{1/3}$ and the split time from the tracks. \textbf{Panel c}, Island depth predicted by the fitted
quartic against the depth measured on the ridge. The prediction grows at a comparable rate but runs ahead of the
measured depth, by a factor of two at $t = 0.56$. Panels d--f, The late split at $\eB = 200$, $\eta = 0.01$.
\textbf{Panel d}, Ridge profiles from $t = 0.9$ to 0.96. \textbf{Panel e}, The fit, which fails once the top has
flattened. \textbf{Panel f}, The flat top from $t = 0.93$ to 0.975. It breaks into several shallow maxima, lattice
ripples here at the $10^{-3}$ level, before the first resolved island appears at $t = 0.959$.}
\label{fig:fit}
\end{figure}

\paragraph{Direct fits.} Supplementary Fig.~\ref{fig:fit} shows the time-resolved fits of the quartic to the ridge
profiles of two reruns with snapshots every 0.005. For the fold at $\eB = 100$, $\eta = 0.05$ the coefficient $a$
rises through zero at $t = 0.538$. The coefficient $b$ falls from 0.012 at $t = 0.50$ to 0.008 at the birth and
changes sign by $t = 0.56$. The fitted coefficients meet the fold condition at $t = 0.546$, by linear interpolation
between snapshots. The critical-point tracks give 0.545 for the first detection and 0.548 for the fitted birth. For
the late split of the flat sheet at $\eB = 200$, $\eta = 0.01$ the fit fails once the top has flattened, at
$t = 0.917$.

\section{\texorpdfstring{The three-dimensional run to $t = 8$}{The three-dimensional run to t = 8}}
\label{sec:3d}
The $\eB = 50$, $\eta = 0.01$ run of Fig.~5a of the main text was continued from $t = 4$ to 8. Its state was embedded,
with unchanged cell sizes, in the larger box $|x|, |y| < 5$, $|z| < 8.25$ ($320\times320\times384$ points). The
energies and the helicity continue across the extension to within 1\%. Between $t = 4$ and 8 the kinetic energy
falls from 52.2 to 17.0, the magnetic energy from 3.19 to 1.72 and the helicity from $-0.0106$ to $-0.0071$. The
ratio $M/|H|$ therefore decreases from 300 to 240. This is consistent with selective decay, although expansion also
lowers it. The departure from axisymmetry decays as well. The fraction of $|\mathbf B|^2$ outside the axisymmetric
mode falls from $6.3\times10^{-3}$ to $6.3\times10^{-4}$. The $m = 2$ fraction falls from $5.6\times10^{-3}$ to
$4.6\times10^{-4}$ and the $m = 1$ fraction from $5.5\times10^{-4}$ to $1.7\times10^{-4}$. These fractions are
sampled inside $r < 2$ and $|z| < 0.8L_z$, with $L_z$ the half-height of the box (Methods). They rise by 30\% at the
extension because the measuring region grows with the box. The original region holds over 99.9\% of the magnetic
energy to $t = 3$ but 76\% at $t = 4$, when the daughters have reached its edge. The region of the larger box holds
99--100\% from $t = 4$ to 8. At $t = 8$ the magnetic axes of the daughters are at $z = 4.8$ and $-5.2$. The 5\% flux
surface stays at least two units inside the box faces, and the neck still carries an island. Supplementary Movie~3
shows this run from $t = 0.1$ to 8.

\renewcommand{\refname}{Supplementary References}
\begingroup\small
\raggedright

\endgroup
\end{document}